\documentclass[aps,prd,showpacs,amsmath,amssymb,twocolumn,superscriptaddress]{revtex4-1}
\usepackage[]{latexsym}
\usepackage[english]{babel}
\usepackage{graphicx}
\usepackage{hyperref}
\usepackage{color}
\def\d{{\rm d}}

\begin{document}

\title{Precision cosmology with Pad\'e rational approximations: theoretical predictions versus observational limits}

\author{Alejandro Aviles}
\email{aviles@ciencias.unam.mx}
\affiliation{Departamento de Matem\'aticas, Cinvestav del Instituto Polit\'ecnico Nacional (IPN), 07360, M\'exico DF, Mexico.}

\author{Alessandro Bravetti}
\email{bravetti@icranet.org}
\affiliation{Instituto de Ciencias Nucleares, Universidad Nacional Aut\'onoma de M\'exico (UNAM), M\'exico, DF 04510, Mexico.}

\author{Salvatore Capozziello}
\email{capozzie@na.infn.it}
\affiliation{Dipartimento di  Fisica, Universit\`a di Napoli ``Federico II", Via Cinthia, I-80126, Napoli, Italy.}
\affiliation{Istituto Nazionale di Fisica Nucleare (INFN), Sezione di Napoli, Via Cinthia, I-80126 Napoli, Italy.}
\affiliation{Gran Sasso Science Institute (INFN), Viale F. Crispi, 7, I-67100, L'Aquila, Italy.}

\author{Orlando Luongo}
\email{orlando.luongo@na.infn.it}
\affiliation{Instituto de Ciencias Nucleares, Universidad Nacional Aut\'onoma de M\'exico (UNAM), M\'exico, DF 04510, Mexico.}
\affiliation{Dipartimento di  Fisica, Universit\`a di Napoli ``Federico II", Via Cinthia, I-80126, Napoli, Italy.}
\affiliation{Istituto Nazionale di Fisica Nucleare (INFN), Sezione di Napoli, Via Cinthia, I-80126 Napoli, Italy.}

\date{\today}

\begin{abstract}
We propose a novel approach for parameterizing the luminosity distance, based on the use of rational ``Pad\'e'' approximations. This new technique extends standard Taylor treatments, overcoming possible convergence issues at high redshifts plaguing standard cosmography. Indeed, we show that Pad\'e expansions enable us to confidently  use data over a larger interval with respect to the usual Taylor series. To show this property in detail, we propose several Pad\'e expansions and we compare these approximations with cosmic data, thus obtaining cosmographic bounds from the observable universe for all cases. In particular, we fit Pad\'e luminosity distances with observational data from different uncorrelated
surveys. We employ union 2.1 supernova data, baryonic acoustic oscillation, Hubble space telescope measurements and differential age data. In so doing, we also demonstrate that the use of Pad\'e approximants can improve the analyses carried out by introducing cosmographic auxiliary variables, i.e. a standard technique usually employed in cosmography in order to overcome the divergence problem. Moreover, for any drawback related to standard cosmography, we emphasize possible resolutions in the framework of Pad\'e approximants. In particular, we investigate how to reduce systematics, how to overcome the degeneracy between cosmological coefficients, how to treat divergences and so forth. As a result, we show that cosmic
bounds are actually refined through the use of Pad\'e treatments and the thus derived best values of the cosmographic parameters show slight departures from the standard cosmological paradigm. Although all our results are perfectly consistent with the $\Lambda$CDM model, evolving dark energy components different from a pure cosmological constant are not definitively ruled out. Finally, we use our outcomes to reconstruct the effective universe equation of state, constraining the dark energy term in a model independent way.
\end{abstract}


\maketitle

\section{Introduction}
\label{Intro}


One of the most challenging  issues of modern cosmology is to describe the positive late time  acceleration~\cite{SNeIa1,SNeIa2,copeland}
through a single self-consistent theoretical scheme.
Indeed, the physical origin of the measured cosmic speed up is not well accounted on theoretical grounds, without invoking the existence of an additional fluid which drives
the universe dynamics, eventually dominating over the other species.
Any viable fluid differs from standard matter by manifesting negative equation of state parameters,
capable of counterbalancing the gravitational attraction at late times~\cite{antig}.
Thus, since no common matter is expected to behave anti-gravitationally, one refers to such a fluid as dark energy.
The simplest candidate for dark energy consists in introducing within Einstein equations a vacuum energy cosmological constant term, namely $\Lambda$~\cite{v1}.
The corresponding paradigm, dubbed the $\Lambda$CDM model, has been probed to be consistent with almost all experimental constraints~\cite{tests},
becoming the standard paradigm in cosmology.
One of the main advantages of $\Lambda$CDM is the remarkably small number of cosmological parameters that it introduces,
which suggests that any modifications of Einstein's gravity reduce to $\Lambda$CDM at small redshift~\cite{freddo}.
However, recent  measurements of the Hubble expansion rate at redshift $z = 2.34$~\cite{Delubac:2014aqe}
and an analysis of linear redshift space distortions~\cite{Macaulay:2013swa} reside outside the $\Lambda$CDM expectations at $2.5\, \sigma$  and  $0.99$ confidence level, respectively.
Due to these facts and to the need of accounting for the ultraviolet modifications of Einstein's gravity,
extensions of general relativity have been proposed so far.
Moreover, the standard cosmological model is plagued by two profound shortcomings.
First, according to observations, it is not clear why matter and $\Lambda$ magnitudes appear to be extremely close to each other, indicating an unexpected coincidence problem~\cite{chomp}.
Second, cosmological bounds on $\Lambda$ indicate a value which differs from quantum field calculations by a factor of 123 orders of magnitude, leading to a severe fine-tuning problem~\cite{problema2}.

Standard cosmology deems that the universe dynamics can be framed assuming that dark energy  evolves as a perfect fluid,
with a varying equation of state, i.e. $\omega(z)\equiv P / \rho$, with total pressure $P$ and density $\rho$. So, in a Friedmann-Robertson-Walker (FRW)
picture, the universe dynamics is depicted through a pressureless matter term, a barotropic evolving dark energy density and a vanishing scalar
curvature, i.e. $\Omega_{k}=0$~\cite{freddo2}.
In lieu of developing a theory which predicts the dark energy fluid, cosmologists often try to reconstruct the universe expansion history,
by parameterizing the equation of state of dark energy~\cite{nexu}.
For example,  polynomial fits, data depending reconstructions and cosmographic
representations~\cite{nexu2} are consolidated manners to reconstruct $\omega(z)$~\cite{capo,cpl,capo2,capo3,capo4,noi}.
All cosmological recontructions are based on inferring the properties of dark energy
 without imposing \emph{a priori} a form for the equation of state.
In fact, any imposition would cause misleading results~\cite{mine1},
as a consequence of the strong degeneracy between cosmological models.
Therefore, it turns out that a reconstruction of $\omega(z)$ should be carried out as much as possible in a model independent manner~\cite{mine2}.
To this regard, a well established method is to develop a model independent parametrization by expanding  $\omega(z)$
into a truncated Taylor series and  fixing the corresponding free parameters through current data~\cite{cpl,altro}.
However, even though Taylor series are widely used to approximate known functions with polynomials around some point, they provide
bad convergence results over a large interval, since $\omega(z)$ is expanded around $z=0$,
while data usually span over intervals larger than the convergence radius~\cite{aggiungiamo,john}.
A more sophisticated technique of approximation, the Pad\'e approximation,
aims to approximate functions by means of a ratio between two polynomials~\cite{padex}.
Pad\'e approximation is usually best suited to approximate diverging functions and functions over a whole interval,
giving a better approximation than the corresponding truncated Taylor series~\cite{padexapprox}.

In~\cite{OrlChri}, Pad\'e approximations have been introduced in the context of cosmography, whereas applications have been discussed and extended in~\cite{Chinese}, but
the authors have focused principally on writing the dark energy equation of state as a Pad\'e function.
In this work we want to propose a new approach to cosmography, based on approximating the luminosity distance by means
of Pad\'e functions, instead of Taylor polynomials.
In this way we expect to have a better match of the model with cosmic data and to overcome possible divergences of the Taylor approach at  $z\gg1$.
Indeed, using the Pad\'e approximation of the luminosity distances, we also show that one can improve the quality of the fits with
respect to  the standard re-parametrizations of the luminosity distances by means of auxiliary variables.
We also propose how to deal numerically with such approximations and how to get the most viable Pad\'e expansions.
As a result, we will obtain a refined statistical analysis of the cosmographic parameters.
A large part of the work will be devoted to
outline the drawbacks and the advantages of this technique and compare it to more standard approaches as Taylor series and the use of auxiliary variables.
Moreover, we also include a discussion about the most adequate Pad\'e types among the wide range of possibilities.
Finally, we obtain a reconstruction of the dark energy equation of state which is only based on
the observational values of the luminosity distance, over the full range for the redshift in which data are given.
In this way, we demonstrate that Pad\'e approximations are actually preferred to fit high redshift cosmic data, thus representing a valid alternative technique to reconstruct
the universe expansion history at late times.

The paper is structured as follows: in Sec.~\ref{model} we highlight the role of cosmography in the description of the present time dynamics of the universe.
In particular, we discuss  connections with the cosmographic series and the FRW metric.
In Sec.~\ref{sec:padeandcosmography} we introduce the Pad\'e formalism
and we focus on the differences between standard Taylor expansions and rational series
in the context of cosmography, giving a qualitative indication that a Pad\'e approximation could be preferred. We also enumerate some issues
related to cosmography in the context of the observable universe. For every problem, we point out possible solutions and we underline
how we treat such troubles in our paper, with particular attention to the Pad\'e formalism. All experimental results have been portrayed in Secs.~\ref{Sect:DataSet} and~\ref{ParameterEstimations}, in which we present
the numerical outcomes derived both from using the Pad\'e technique
and standard cosmographic approach. In Sec.~\ref{applicationsPade} we give an application of the Pad\'e recipe, that is,
we use the Pad\'e technique to estimate the free parameters of some known models. In Sec.~\ref{universeEoS}, we discuss the consequences on the equation of state for the universe which can be inferred from our numerical
outcomes derived by the use of Pad\'e approximants. Moreover, in Sec.~\ref{consequencespade} we discuss our numerical outcomes and we interpret the bounds obtained. Finally, the last section, Sec.~\ref{conclusions}, is devoted to
conclusions and perspectives of our approach.


\section{The role of cosmography in \emph{precision cosmology}}
\label{model}

In this section we briefly introduce the role of cosmography and its standard-usage techniques to fix cosmographic
constraints on the observable universe. The great advantage of the cosmographic method is that it permits
one to bound present time cosmology
without having to assume any particular model for the evolution of dark energy with time.
The \emph{cosmographic method} stands for a coarse grained model
independent technique to gather viable limits on the universe expansion history at late times, provided the cosmological
principle is valid~\cite{mine1,mine2,cosmus}. The corresponding requirements demanded by cosmography are
homogeneity and isotropy with spatial curvature somehow fixed.
Common assumptions on the cosmological puzzle provide a whole energy budget dominated by $\Lambda$,
(or by some sort of dark energy density), with cold dark matter in second place and baryons as a small fraction only.
Spatial curvature in case of time-independent dark energy density is actually constrained to be negligible.
However, for evolving dark energy contributions, observations are not so restrictive~\cite{rat}.
More details will be given later, as we treat the degeneracy between scalar curvature and variation of the acceleration.
From now on, having fixed the spatial curvature $\Omega_{k}$ to be zero, all cosmological observables can be expanded around present time.
Moreover, comparing such expansions to cosmological data allows one to fix bounds on the evolution of each
variable under exam. This strategy  matches cosmological observations with theoretical expectations.
By doing so, one gets numerical outcomes which do not
depend on the particular choice of the cosmological model, since only Taylor expansions are compared with data.
Indeed, cosmography relates observations and theoretical predictions, and it is able to alleviate the degeneracy
among cosmological models. Cosmography is therefore able to distinguish between models that are compatible with
cosmographic predictions and models that have to be discarded, since they do not fit the cosmographic limits.

Hence, according to the cosmological principle, we assume the universe to be described by a Friedmann-Robertson-Walker (FRW) metric, i.e.
\begin{equation}\label{frw}
 \d s^2=\d t^2-a(t)^2\left(\d r^2+r^2\d\Omega^2\right)\,,
\end{equation}
where we use the notation $\d\Omega^2\equiv \d\theta^2+\sin^2\theta \d\phi^2$.

As a first example of cosmographic expansions, we determine the scale factor $a(t)$ as a Taylor series~\cite{Weinberg2008} around present time $t_0$. We have
\begin{eqnarray}\label{serie1a}
a(t)  & \sim & a(t_0)+ a'(t_0) \Delta t + \frac{a''(t_0)}{2} \Delta t^2+
\frac{a'''(t_0)}{6} \Delta t^3 +\nonumber\\
&+&   \frac{a^{(iv)}(t_0)}{24} \Delta t^4 +\frac{a^{(v)}(t_0)}{120} \Delta t^5+\ldots\,,
\end{eqnarray}
which recovers signal causality if one assumes $\Delta t\equiv t-t_0>0$. From the above expansion of $a(t)$, one  defines
\begin{subequations}\label{CSdef}
\begin{align}
H \equiv \frac{1}{a} \frac{\d a}{\d t}\,,\\
q \equiv -\frac{1}{a H^2} \frac{\d^2a}{\d t^2}\,,\\
j  \equiv \frac{1}{a H^3} \frac{\d^3a}{\d t^3}\,,\\
s \equiv \frac{1}{a H^4} \frac{\d^4a}{\d t^4}\,,\\
l  \equiv \frac{1}{a H^5} \frac{\d^5a}{\d t^5}\,.
\end{align}
\end{subequations}
Such functions are, by construction,  model independent quantities, i.e. they do not depend on the form of the dark energy fluid,
since they can be directly bounded by observations.
They are known in the literature as the Hubble rate ($H$), the acceleration parameter ($q$), the jerk parameter ($j$),
the snap parameter ($s$) and the lerk parameter ($l$)~\cite{starobinskievisser}. Once such functions are fixed at present time,
they are referred to as the \emph{cosmographic series} (CS). This is the set of coefficients usually derived in cosmography from observations.

Rewriting $a(t)$ in terms of the CS gives
\begin{eqnarray}\label{serie1a2}
a(t)  & \sim & 1+  H_0 \Delta t - \frac{q_0}{2}  H_0^2\Delta t^2+\nonumber \\
&+&
\frac{j_0}{6} H_0^3 \Delta t^3 +   \frac{s_0}{24}  H_0^4\Delta t^4 +\frac{l_0}{120}  H_0^5\Delta t^5\ldots\ ,
\end{eqnarray}
where we have normalized the scale factor to $a(t_0) = 1$. By rewriting Eq.~($\ref{serie1a}$) as in Eq.~($\ref{serie1a2}$), one can read out the meaning of each parameter.
In fact, each term of the CS displays a remarkable dynamical meaning.
In particular, the
snap and lerk parameters determine the shape of Hubble's flow at higher redshift regimes. The Hubble parameter must be
positive, in order to allow the universe to expand and finally $q$ and $j$ fix kinematic properties at lower redshift domains. Indeed,
the value of $q$ at a given time specifies whether the universe is accelerating or decelerating and also provides some hints on
the cosmological fluid responsible for the dynamics.
Let us focus on $q$ first. We can distinguish three cases, splitting the physical interval of viability for $q_0$:
\begin{enumerate}
\item
    $\underline{q_0>0}$, shows an expanding universe which undergoes a deceleration phase.
    This is the case of either a matter dominated universe or any pressureless barotropic fluid.
    Observations do not favor $q_{0}>0$ at present times, which however appears relevant for early time cosmology,
    where dark energy did not dominate over matter.

\item
    $\underline{-1<q_0<0}$, represents an expanding universe which is currently speeding up. This actually represents
    the case of our universe.    The universe is thought to be dominated by some sort of anti-gravitational fluid, as stressed in Sec.~\ref{Intro}.
    In turn, cosmography confirms such characteristics,    without postulating any particular form of dark energy evolution.

\item
    $\underline{q_0=-1}$,  indicates that all the whole cosmological energy budget is dominated by a de Sitter fluid, i.e. a cosmic component with
    constant energy density which does not evolve as the universe expands. This is the case of Inflation at the very early
    universe. However at present-time this value is ruled-out by observations.

\end{enumerate}

Besides, the variation of the acceleration provides a way to understand whether the universe passes or not through a
deceleration phase. Precisely, the variation of acceleration, i.e. $\d q/\d z$, is related to $j$ as
\begin{equation}\label{jh35kdj}
\frac{\d q}{\d z}=\frac{j-2q^2-q}{1+z}\,.
\end{equation}
At present time, we therefore have $j_0=\frac{\d q}{dz}|_{0}+2q_0^2+q_0$ and since we expect $-1\leq q_0<-1/2$, we get that
$2q_0^2+q_0>0$. Thus if $q_0 < -1/2$, then $j_0$ is linked to the sign of the variation of $q$. We will confirm from observations that it actually lies in the interval $q_0<-1/2$.

Accordingly we can determine three cases:

\begin{enumerate}
 \item   $\underline{j_0<0}$ the universe does not show any departure from the present time accelerated phase.
  This would indicate that dark energy influences early times dynamics, without any changes throughout the universe evolution.
  Even though this may be a possible scenario, observations seem to indicate that this does not occur and it
  is difficult to admit that the acceleration parameter does not change its sign as the universe expands.

  \item  $\underline{j_0=0}$ indicates that the acceleration parameter smoothly tends to a precise value, without changing
  its behavior as $z\rightarrow\infty$. No theoretical considerations may discard or support this hypothesis, although
  observations definitively show that a model compatible with zero jerk parameter badly fits current cosmological data.

\item
 $\underline{j_0>0}$ implies that the universe acceleration started at a precise time during the  evolution.
 Usually, one refers to the corresponding redshift as the \emph{transition redshift}~\cite{transition}, at which dark energy effects
 actually become significative. As a consequence, $j_0>0$ indicates the presence of a further cosmological \emph{resource}. By a direct measurement of the transition redshift $z_{tr}$, one would get relevant constraints on
 the dark energy equation of state.
 It turns out that the sign of $j_0$ corresponds to a change of slope of universe's dynamics. Rephrasing it differently, a positive $j_0$ definitively forecasts that the acceleration parameter should change sign at $z>z_{tr}$.
\end{enumerate}

A useful trick of cosmography is to re-scale the CS by means of the Hubble rate. In other words, it is possible to
demonstrate that if one takes into account $n$ cosmographic coefficients, only $n-1$ are really independent.
 From definitions~(\ref{CSdef}), one can write
\begin{subequations}\label{Hpunto}
\begin{align}
\dot{H} =& -H^2 (1 + q)\,,\\
\ddot{H} =& H^3 (j + 3q + 2)\,,\\
\dddot{H} =& H^4 \left [ s - 4j - 3q (q + 4) - 6 \right]\,,\\
\ddddot{H} =& H^5 \left [ l - 5s + 10 (q + 2) j + 30 (q + 2) q + 24\right ]\,,
\end{align}
\end{subequations}
\noindent and we immediately see the correspondence between derivatives of the Hubble parameter and the CS
(note in particular the degeneracy, due to the fact that all these expressions are multiplied by $ H$).

As a consequence of the above discussion, one can choose a particular set of observable quantities and, expanding it,
as well as the scale factor, it is possible to infer viable limits on the parameters. To better illustrate this statement, by means
of Eqs.~(\ref{Hpunto}), one can infer the numerical values of the CS using the well-known luminosity distance.

In fact, keeping in mind the definition of the cosmological redshift $z$ in terms of the cosmic time $t$, that is
\begin{equation}\label{cosmoz}
\frac{\d { z} }{(1+z)}=-H(z)\d t\,,
\end{equation}
then the luminosity distance in flat space can be expressed as
\begin{equation}\label{defDL}
    d_L  =   (1+z)\chi(z),
\end{equation}
where
\begin{equation}\label{kjjgjhg}
\chi(z) =\int_{0}^{z}{\frac{dz'}{H(z')}}\,,
\end{equation}
is the comoving distance traveled by a photon from redshift $z$ to us, at $z=0$. The $d_L$ can be written as
\begin{eqnarray}\label{dLTaylor}
d_L = z\,d_H(H_0) \tilde{d}_L(z;q_0,j_0,\dots)\,,
\end{eqnarray}
where
\begin{equation}\label{hdhfkjd}
d_H(H_0)= \frac{1}{H_0}\,,
\end{equation}
and
\begin{eqnarray}\label{jhgkfh}
\tilde{d}_L(z;q_0,j_0,\dots) &=& 1 +\frac{1 - q_0}{2}z+\\
&-& \frac{1-q_0(1+3q_{0}) +j_0 }{6}z^2 + \mathcal{O}(z)\,.\nonumber
\end{eqnarray}
Further expansions up to order five in $z$ that will be used in this work are reported in the Appendix~\ref{appA}. Here, for brevity we reported in Eq. (\ref{jhgkfh}) the expansion up to the second order in $z$.

It is worth noticing that Eq.~(\ref{dLTaylor}) is general and applies to any cosmological model, provided it
is based on a flat FRW metric. Thus, by directly fitting the cosmological data for $d_L$, one gets physical bounds on $q_0$,
$j_0$, $s_0$ and $l_0$ \emph{for any cosmological model} (see~\cite{mine1,mine2}).

The above description, based on common Taylor expansions, represents only one of the possible approximations that one may use for the luminosity distance.
It may be argued that such an approximation does not provide adequate convergence for high redshift data.
Thus, our aim is to propose possible extensions of the standard Taylor treatment, i.e. Pad\'e approximants, that could accurately resolve the issues of standard cosmography.

In the next section we will present a different approximation for the luminosity distance, given by rational Pad\'e functions instead of Taylor polynomials,
we will analyze the relationship with the usual Taylor expansion
and argue that the rational approximation may be preferred from a theoretical point of view.
Later, in Secs.~\ref{Sect:DataSet} and~\ref{ParameterEstimations}
we will also perform the numerical comparison with observational data, in order to show that one can get improved results from this novel approach.


\section{Pad\'e approximations in the context of cosmography}\label{sec:padeandcosmography}

In this section we introduce the concept of Pad\'e approximants and we describe the applications of Pad\'e treatment to cosmography.

To do so, let us define the $(n,m)$ Pad\'e approximant of a generic function $f(z)$, which is given by the rational function
\begin{equation}\label{padedef}
P_{n m}(z)  = \frac{a_0+a_1\,z+\dots+a_n\,z^n}{1+b_1\,z+\dots+b_m\,z^m}\,,
\end{equation}
with degree $n\geq0$ (numerator) and $m\geq0$ (denominator) that agrees with $f(z)$ and its derivatives at $z=0$ to the highest possible order,  i.e. such that $P_{n m}(0)=f(0), P_{n m}'(0)=f'(0),\ldots,  P_{n m}^{m+n}(0)=f^{m+n}(0)$~\cite{Padebook}. Pad\'e approximants for given $n$ and $m$ are unique up to an overall multiplicative constant. As a consequence, the first constant in the denominator is usually set to one, in order to face this scaling freedom. Hereafter, we follow this standard notation and indicate as $P_{nm}$ the Pad\'e approximant of degree $n$ at the numerator and $m$ in the denominator. As we see, in cosmology one may use direct data through the distance modulus $\mu$ of
different astronomical objects, such as e.g. supernovae. In the usual applications of cosmography, the luminosity distance
$d_L$, which enters $\mu$, is assumed to be a (truncated) Taylor series with respect to $z$ around present time $z=0$.
A problem with such a procedure occurs when one uses data out of the interval $z\geq1$. In fact, due to the divergence at high redshifts of the Taylor polynomials, this can possibly give non-accurate numerical results \cite{OrlChri}.
Consequently, data taken over $z>1$ are quite unlikely to accurately fit Taylor series. Pad\'e approximants can resolve this issue. In fact let us consider the general situation when one has to reconstruct a function, supposing to know the values of such a function,  taken in the two limits where the independent variable is very small and very large respectively. Hence, let us consider two different approximate expansions of $d_L$. The first for small values of $z$ (around $z=0$), the second for large values of $z$ (around $1/z=0$).
The two approximations can be written as
\begin{subequations}\label{sus}
\begin{align}
d_L^0&\sim f_0+f'\,z+\frac{f''}{2}\,z^2\\
d_L^{\infty}&\sim g_0+g'\,\frac{1}{z}+\frac{g''}{2}\,\frac{1}{z^2}\,.
\end{align}
\end{subequations}
In this way, provided we construct a function that behaves as $d_L^0$ when $z \sim 0$ and as $d_L^\infty$ as $z \sim \infty$,
we are sure that in both limits such a function remains finite (when $z\sim 0$ and $z\sim \infty$ respectively). Given such a property, the most natural function able to interpolate our data between those two limits is naturally given by a \emph{rational function} of $z$. Pad\'e approximants are therefore adequate candidates to carry on this technique. In the next subsection we describe some problems associated to cosmography and to the Pad\'e formalism. Later, we also propose feasible solutions that we will adopt throughout this work.

\subsection{Pad\'e treatment to overcome cosmographic drawbacks}

We introduce this subsection to give a general discussion about several drawbacks plaguing the standard cosmographic approach.
For every single problem, we describe the techniques of solutions in the framework of Pad\'e approximations, showing how we treat Pad\'e approximants in order to improve the cosmographic analysis.

\begin{description}
  \item[Degeneracy between coefficients]
  Each cosmographic coefficient may be related to $H$, as previously shown.
  This somehow provides that the whole list of independent parameters is really limited to $q_0,j_0,s_0,l_0,\ldots$. However,
  one can think of measuring $ H_0$ through cosmography in any case, assuming $ H_0$ to be a cosmographic coefficient, without loss of generality.
      The problem of degeneracy unfortunately leads to the impossibility of estimating $ H_0$ alone by using measurements of the distance modulus,
      \begin{equation}
       \mu(z) = 5\log_{10}\left(\frac{d_L(z)}{\text{Mpc}}\right) + 25\,,
      \end{equation}
      in the case of supernova observations, as we will see later. From Eq.~(\ref{dLTaylor}) it follows that $d_L$ can be factorized into two pieces: $d_H$ and $\tilde{d}_L$.
      Since $d_H \equiv H_0^{-1}$, therefore it
      depends only on $H_0$, thus  becoming
       an additive constant in $\mu(z)$ which cannot be estimated, its only effect is to act as a lever to
       the logarithm of $\tilde{d}_L(z;q_0,j_0,\dots$).

       In other words, $H_0$ degenerates
      with the rest of the parameters. To alleviate such problem, we here make use of two different data sets, together with supernovae, i.e.
      the Hubble measurements and the Hubble telescope measure. In this way we employ direct measures of $ H_0$, thus reducing
      the errors associated to the degeneracy between cosmographic parameters.

  \item[Degeneracy with scalar curvature] Spatial curvature of the FRW model
  enters the luminosity distance, since the metric directly depends on it. Thus, geodesics of photons correspondingly change due to its value.
  Therefore, any expansion of $d_L$ depends on $\Omega_k$ as well, degenerating the values of the CS with respect to $\Omega_k$.
  The jerk parameter is deeply influenced by the value of the scalar curvature and degenerates with it. In our work, we
  overcome this problem through geometrical bounds on $\Omega_k$, determined by early time observations.
  According to recent measurements, the universe is considered to be spatially flat and any possible
  small deviations will not influence the simple case $\Omega_k=0$.
  This is the case we hereafter adopt, except for the last part in which we extensively investigate the role of $\Omega_k$ in the framework of the $\Lambda$CDM model.

  \item[Dependence on the cosmological priors]
  The choice of the cosmological priors may influence the numerical outcomes derived from our analyses.
  This turns out to be dangerous in order to determine the signs of the cosmographic coefficients.
  However, to alleviate this problem we may easily enlarge all the cosmological priors, showing that within convergence ranges
  the CS are fairly well constrained.
  The corresponding problem would indicate possible departures from convergence limits, if ranges are outside the theoretical expectations.
  Hence, we found a compromise for each cosmological interval, and we report the whole list of numerical priors in Table~\ref{tab:priors}.

  \item[Systematics due to truncated series]
  Slower convergence in the best fit algorithm may be induced by choosing truncated series at a precise order,
  while systematics in measurements occur, on the contrary, if series are expanded up to a certain order.
  In other words, introducing additional terms would decrease convergence accuracy,
  although lower orders may badly influence the analysis itself. To alleviate this problem, we will constrain the parameters
  through different orders of broadening samples.  In this way, different orders will be analyzed and we will show no significative departures from our truncated series order.

      \item[Dependence on the Friedmann equations]
      Dark energy is thought to be responsible of the present time acceleration.
      However, cosmography is able to describe the current universe speeding up without the need of postulating a precise dark energy fluid \emph{a priori}.
      This statement is clearly true only if a really barotropic fluid is responsible for the dark energy effects.

      In case there is no significative deviations from a constant equation of state for pressureless matter and dark energy is provided by
      some modification of gravitation, cosmography should be adjusted consequently.

      This leads to the implicit choice of assuming general relativity as the specific paradigm to get constraints on the cosmographic observables.
      One may therefore inquire to what extent cosmography is really independent of the Friedmann equations.
      Rephrasing it differently, to reveal the correct cosmological model we do not fix further assumptions,
      e.g. geometrical constraints, Lorentz invariance violations, and so forth, since we circumscribe our analysis to general relativity only.
      Any possible deviations from the standard approach would need additional theoretical  bounds and the corresponding
      CS should be adjusted accordingly.
      However, this problem does not occur in this work and we {can} impose limits without the need of particular assumptions at
      the beginning of every analysis.

  \item[Convergence]  The convergence problem probably represents the most spinous issue of cosmography.
  As we have previously stated, the problem of truncated series is intimately
  intertwined to the order chosen for determining the particular Taylor expansion under exam.
  Unfortunately, almost all cosmological data sets exceed the bound $z\simeq 0$,
  which represents the value around which one expands $d_L$ into a series.
  In principle, all Taylor series are expected to diverge when $z\gg 1$, as a consequence of the fact that they are polynomials.
  Thus, finite truncations get problems to adapt to data taken at $z\gg1$, leading to possible misleading outcomes.
  For example, this often provides additional systematic errors because it is probable that the increase of bad convergence may affect  numerical results.
  Here, we improve accuracy by adopting the union 2.1 supernovae data set and the two additional surveys based on measurements of $H(z)$,
  i.e. direct Hubble measures and Hubble space telescope measurements.
  Combining these data together naturally eases the issue of systematics,
  whereas to overcome finite truncation problems we manage to develop the so called Pad\'e approximation for different orders.
  By construction, since Pad\'e approximants represent a powerful technique to approximate functions by means of ratios of polynomials,
  one easily alleviates convergence problems for $z\gg1$.
  As such, we expect Pad\'e approximants can better approximate the luminosity distance with respect to standard Taylor treatments,
  especially when high redshift data sets are employed in the analysis.

  On the other hand, in order to overcome the problem of divergence, precision cosmology employs the use of several re-parametrizations of the redshift $z$,
  in terms
  of \emph{auxiliary variables}  ($\mathcal Z_{new}$), which enlarge the convergence radius of the Taylor expansion to a sphere of radius $\mathcal Z_{new}<1$.
  Rephrasing it differently, supposing that  data lie within $z\in[0,\infty)$, any auxiliary variable restricts the interval in a more stringent (non-divergent) range.
  A prototype of such an approach is for example given by $y_1 = \frac{z}{1+z}$ (see e.g.~\cite{Cattoen:2007sk}),
  whose limits in the past universe (i.e. $z \in [0,\infty)$) read $y_1\in [0,1]$, while in the future (i.e. $z \in [-1,0]$) read  $y_1\in (-\infty,0]$.
  The construction of any auxiliary variable should satisfy some additional requirements.
  It must be feasible to invert it, passing from the redshift $z$ to it, being one to one invertible.
  Moreover, it should not diverge for any values of the redshift $z$
  (in this sense, $y_1$ suffers from a divergence problem at future times).
  Finally, any parametrization needs to behave smoothly as the universe expands, without showing any critical points.

  In this work, we also compare Pad\'e expansions with the auxiliary variables proposed in the literature, namely $y_1$,
  already cited, and $y_4=\arctan z$, which has been introduced in~\cite{mine2}.
  The variable $y_4$ improves $y_1$, since it has been constructed by following the above mentioned recipe to build up \emph{ad hoc} auxiliary variables. One of our results is that auxiliary variables, albeit being well consolidated tricks for reducing the convergence problem,
  behave worse than Pad\'e approximations.
  This is probably due to the unknown form of the correct $\mathcal Z_{new}$, which is not known \emph{a priori}.
  To do so, we describe in detail differences between our new technique of cosmographic  investigations which uses Pad\'e approximations
  versus standard approaches which make use of auxiliary variables,
  showing that the convergence problem may be definitively healed through the use of rational approximations, instead of constructing auxiliary variables.
\end{description}

In the next subsection we demonstrate with the help of exactly soluble models that Pad\'e approximations
indeed improve the accuracy in approximating the luminosity distance. We stress the fact that this property is more significative as data span over larger intervals of $z$, i.e. $z\gg1$.

\subsection{Taylor versus Pad\'e for exact cosmological models}
\label{comparisonLCDM}

In this section, we give a qualitative representation of the improvements that one gets by performing a Pad\'e approximation  of the luminosity distance.
To do so, we plot $d_L$ for known models and compare it with the numerical behavior of different Taylor and Pad\'e approximations over a large range of $z$.
As two significative examples, we work with the $\Lambda$CDM and $\omega$CDM models for our elucidative purposes~\cite{galaxyluongo}.
Afterwards, we infer a theoretical method to focus on viable Pad\'e approximations, by treating powers $n$ and $m$, without comparing with particular models.

\begin{figure*} [ht]
\begin{center}
\includegraphics[width=3.5in]{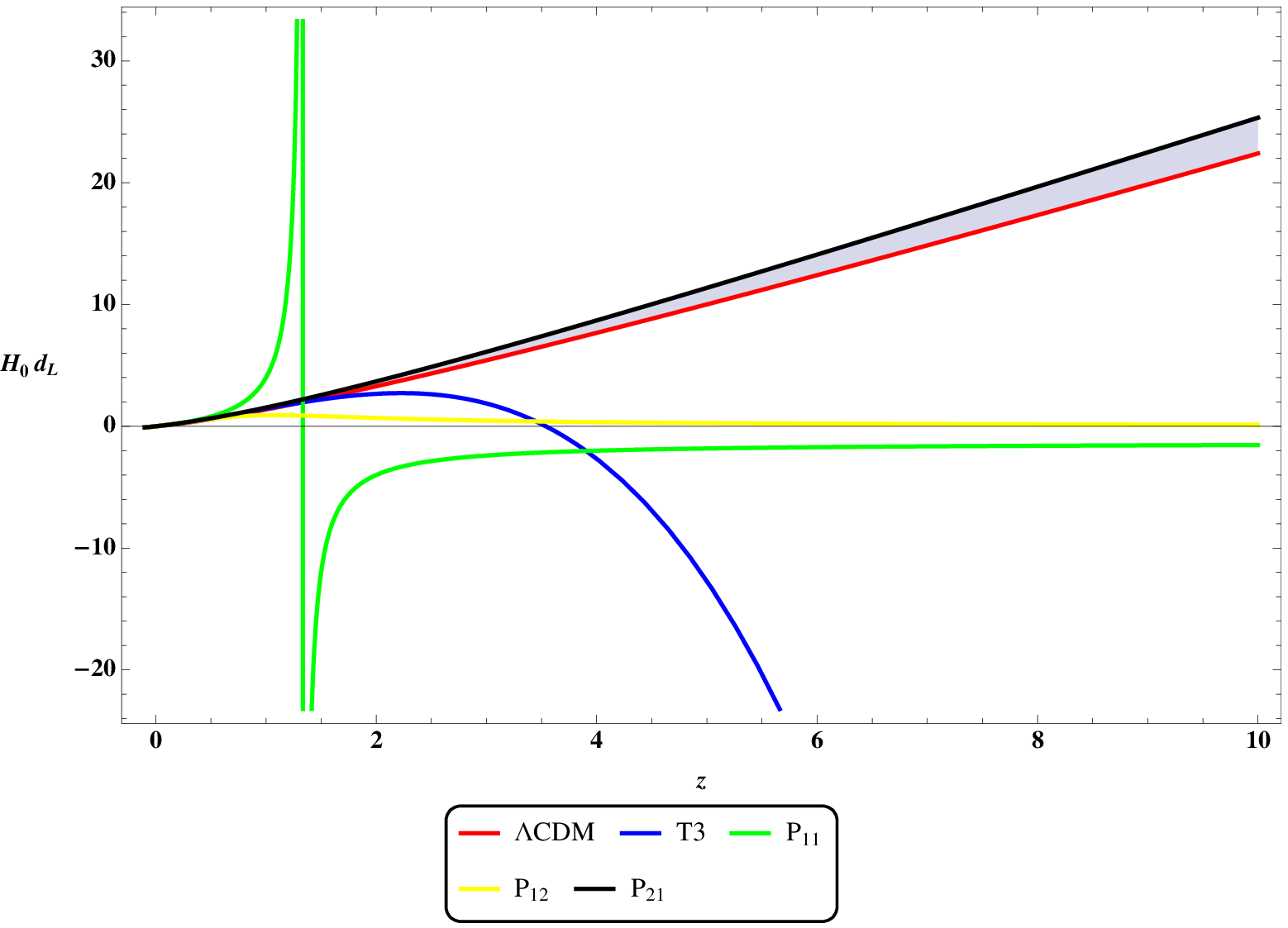}
\includegraphics[width=3.5in]{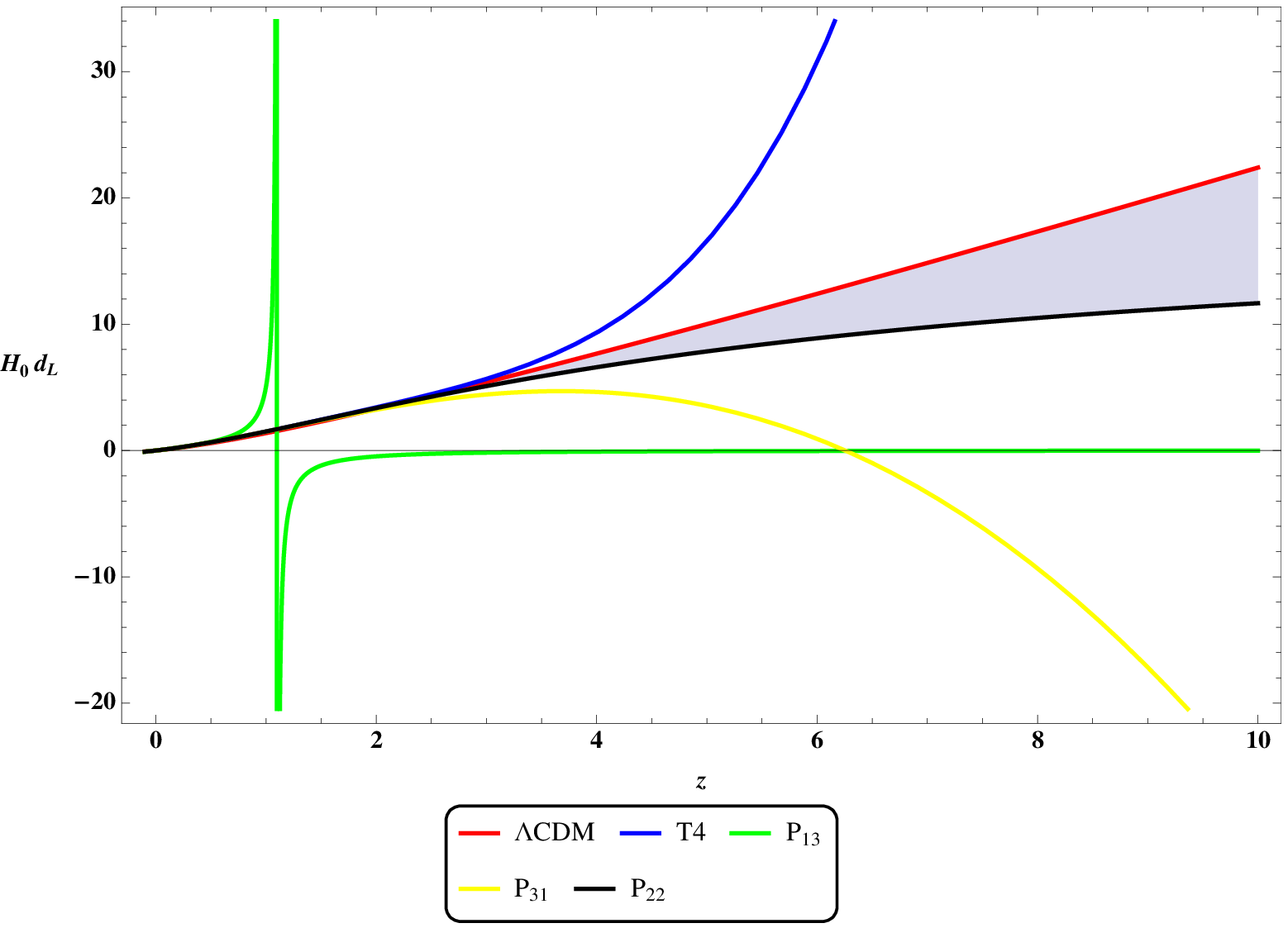}
\includegraphics[width=3.5in]{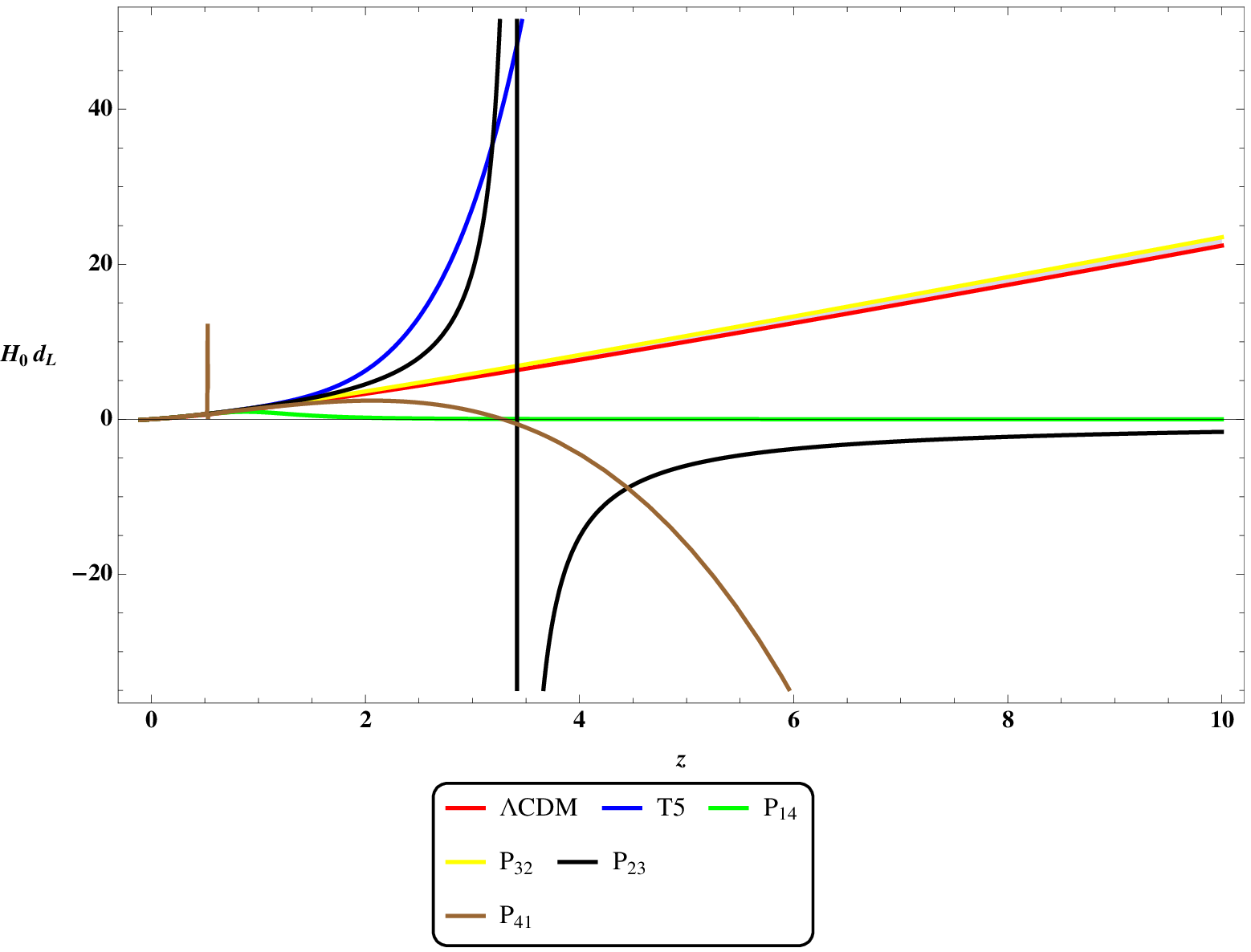}
{\small \caption{Analytical curves for the luminosity distance of the $\Lambda$CDM model (in red) compared
to its Taylor and Pad\'e approximations. As we see, the Taylor polynomials $T3$, $T4$ and $T5$ tends to quickly diverge outside the region $z\leq2$.
At the same time, not all the Pad\'e approximants give good approximations of this model. For example, $P_{11}$, $P_{13}$ and $P_{23}$ give spurious singularities
when used to approximate the $\Lambda$CDM model. We will see how to avoid this problem in the numerical analysis. On the contrary, $P_{21}$, $P_{22}$ and $P_{32}$
give excellent approximations to the $d_{L}$ derived from assuming $\Lambda$CDM.
Additional comments have been reported in the text.}
\label{LCDMDL}}
\end{center}
\end{figure*}

\textbf{The $\Lambda$CDM model:}
The Hubble parameter for the $\Lambda$CDM model reads
\begin{equation}
H(z)=\sqrt{\Omega_m(1+z)^3+\Omega_\Lambda}\,,
\end{equation}
where $\Omega_\Lambda=1-\Omega_{m}$, representing the dark energy density, expressed in terms of a pure cosmological constant. Performing a numerical integration of $d_L$~(\ref{defDL}),
we can plot such  function over the interval of interest, which is arbitrarily fixed inside $z\in[0,10]$. Besides, we can also compute different
Taylor and Pad\'e approximations for this function, and graph all the results, to show that the
approximation is generally improved with the use of rational functions. In Fig.~\ref{LCDMDL},
we present the plots of the exact $\Lambda$CDM luminosity distance,
compared with its approximations obtained using a Taylor polynomial and Pad\'e functions,
for different orders of approximation. In particular, the Taylor polynomial of degree three is plotted
together with the Pad\'e approximants of degree $(1,1)$, $(1,2)$ and $(2,1)$,
the polynomial of fourth degree together with the Pad\'e approximants $(1,3)$, $(3,1)$ and $(2,2)$
and finally the fifth order Taylor polynomial is compared with the Pad\'e functions $(1,4)$, $(3,2)$, $(2,3)$ and $(4,1)$.
We remind that e.g. the Taylor polynomials of third degree and the Pad\'e approximants $(1,2)$ and $(2,1)$
have the same number of free parameters and they agree by definition up to the third order of derivatives at present time.
The same holds to higher orders of both Taylor and Pad\'e approximations. Therefore, the situation described by the Taylor and Pad\'e approximants can also be seen as having
two different models which give approximately the same values for the CS parameters,
albeit providing different evolutions over the whole interval considered.
As one can immediately notice from all plots in Fig.~\ref{LCDMDL}, Taylor
approximations (in blue in the figures) are really accurate until $z$ stays small, whereas they rapidly diverge from the exact curve (in red) as $z>2$.
On the contrary, we can see from the first plot that the rational approximant  $P_{21}$ keeps very close to the exact function over the complete interval analyzed.
Moreover, as we see in the second and third plots, the situation is the same as we increase the order of the approximants.
In fact,  the Pad\'e functions $P_{22}$ and $P_{32}$ fairly approximate the exact $\Lambda$CDM luminosity distance over all the interval considered.
In particular, we remark also that the correctness of the approximation not necessarily increases by increasing the order of the approximants
(as expected, since all the possible Pad\'e functions have completely different behaviors, depending on the degrees of numerator and denominator)
and that $P_{21}$, $P_{22}$ and $P_{32}$ seem to be the best approximations, within the ones considered, giving excellent results.

As a good check for our conclusions, we repeat such considerations by using a different model, i.e. the $\omega$CDM model, probably representing the first step beyond the $\Lambda$CDM model.

\textbf{The $\omega$CDM model:}
The Hubble parameter resulting from the $\omega$CDM model reads
\begin{equation}
H(z)=\sqrt{\Omega_m(1+z)^3+\Omega_Q\,(1+z)^{3(1+\omega)}}\,,
\end{equation}
where $\Omega_Q=1-\Omega_{m}$ and $\omega$ is a
free parameter of the model that lies in the interval $\omega\in (-1,-0.8)$. The exact integral of $d_L$ involves here a hypergeometric function.
We plot it over the interval of interest, which is again $z\in[0,10]$. Besides, we can also compute different Taylor and Pad\'e approximations for this function, plotting all the results, to show that the
approximation is generally improved with the use of rational functions as well as in the $\Lambda$CDM case.

\begin{figure*} [ht]
\begin{center}
\includegraphics[width=3.5in]{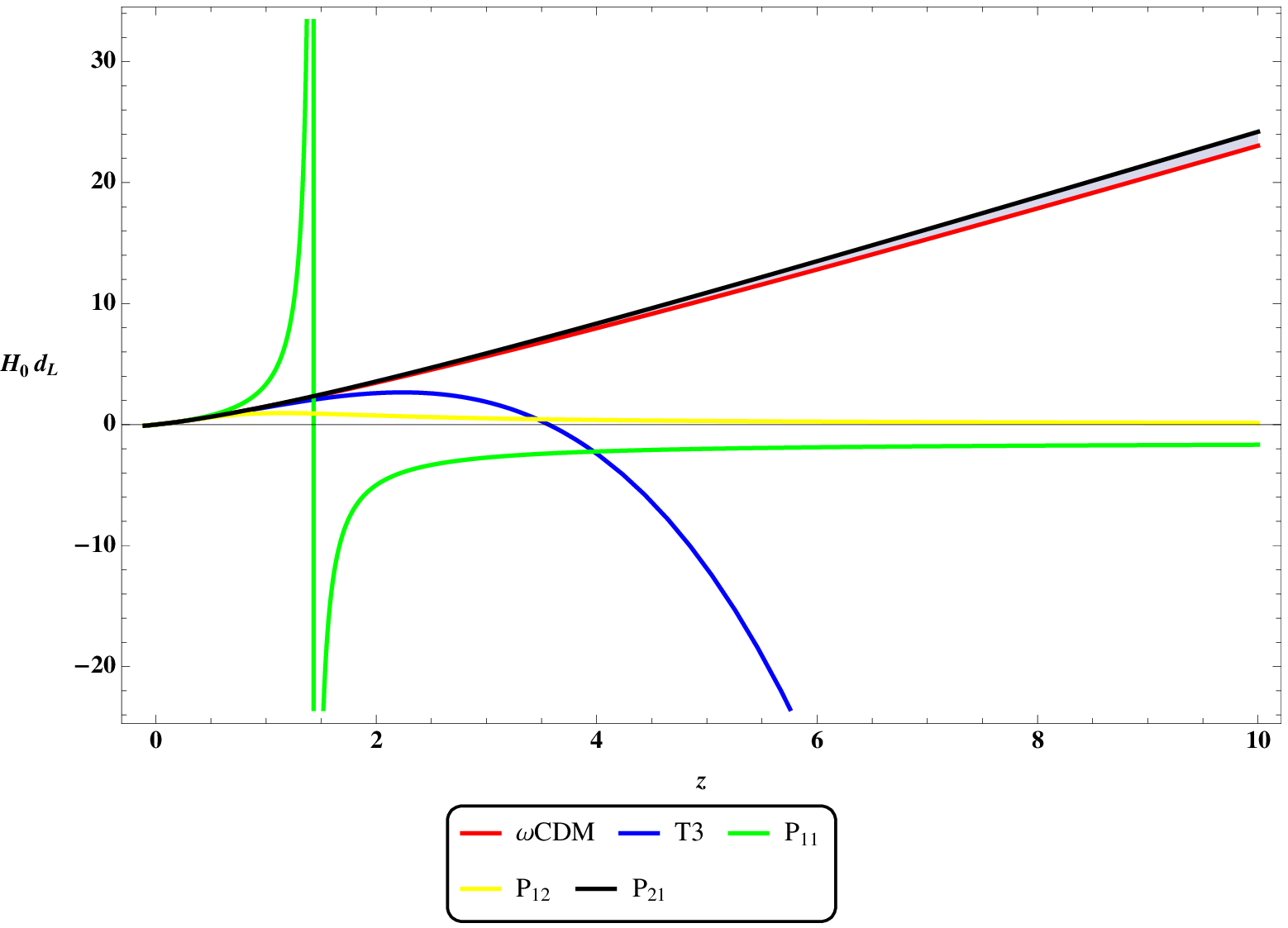}
\includegraphics[width=3.5in]{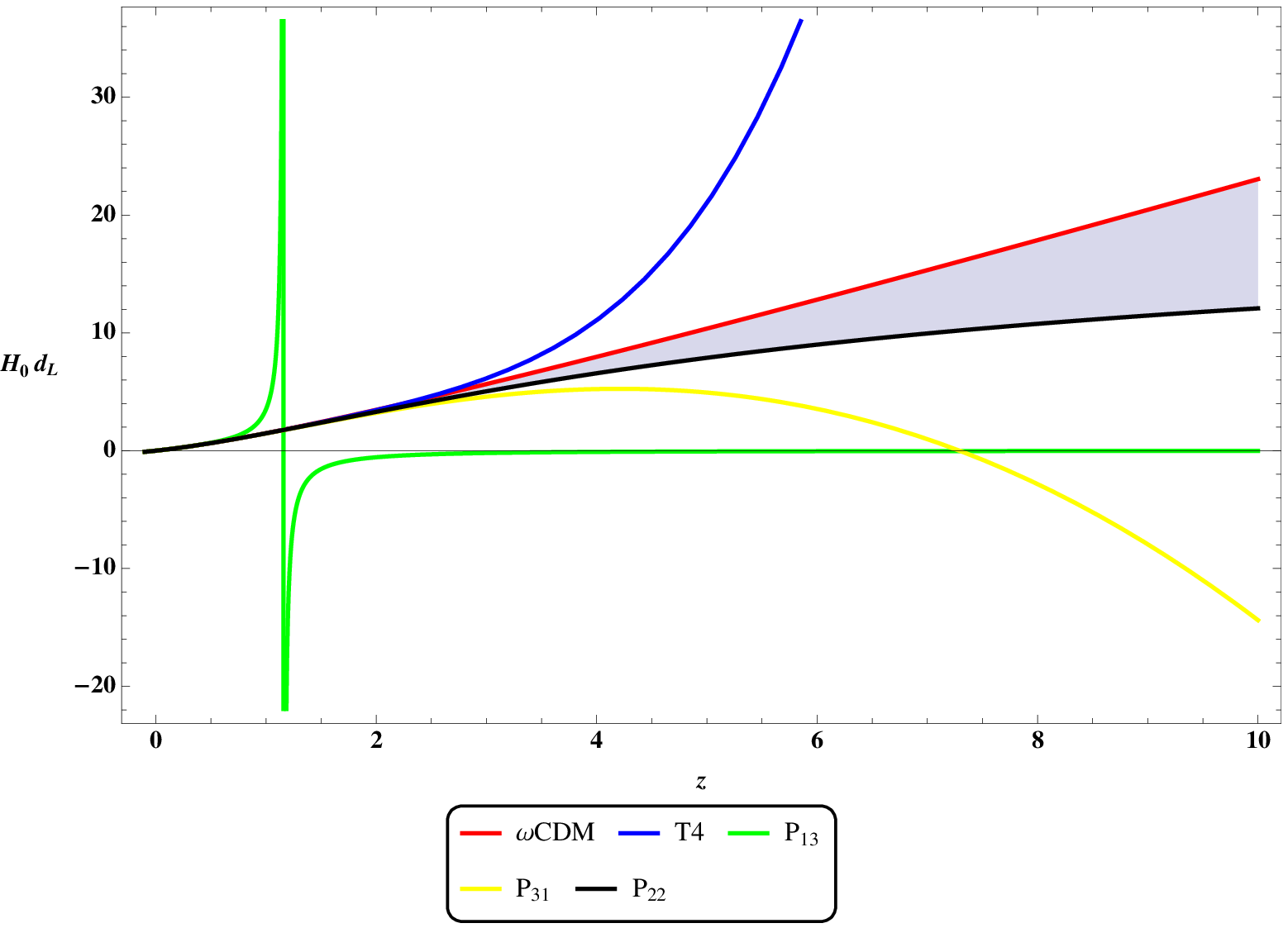}
\includegraphics[width=3.5in]{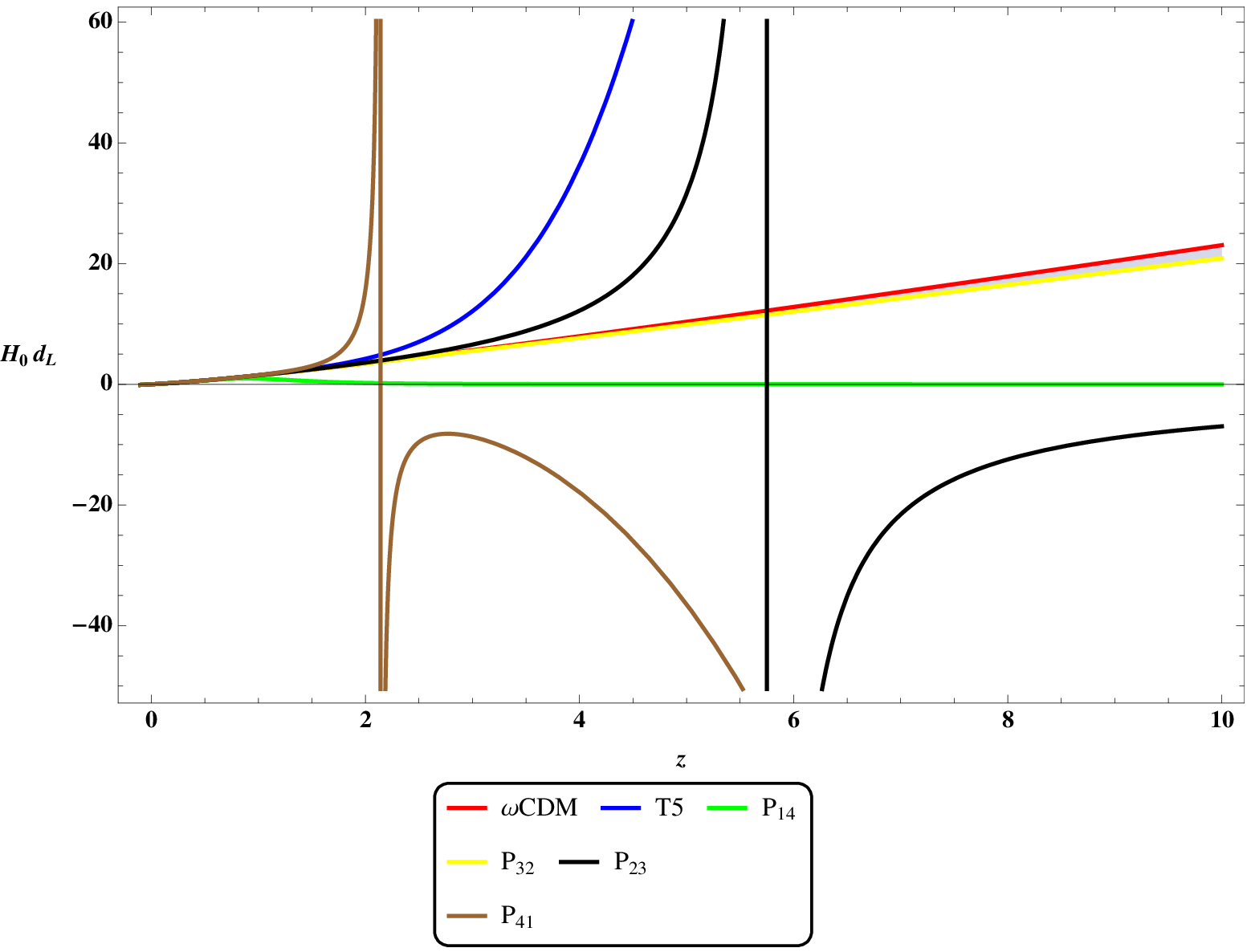}
{\small \caption{Analytical curves for the luminosity distance of the $\omega$CDM model (in red) compared
to its Taylor and Pad\'e approximations. As we see, the Taylor polynomials $T3$, $T4$ and $T5$ tends to quickly diverge outside the region $z\leq2$.
At the same time, not all the Pad\'e approximants give good approximations of this model. For example, $P_{11}$, $P_{13}$, $P_{23}$ and $P_{41}$ give spurious singularities
when used to approximate the $\omega$CDM model. We will see how to avoid this problem before our numerical analysis. On the contrary, $P_{21}$, $P_{22}$ and $P_{32}$
give excellent approximations to the $d_{L}$ derived from assuming $\omega$CDM.
More comments in the text.}
\label{wCDMDL}}
\end{center}
\end{figure*}

Moreover, all the comments presented for the $\Lambda$CDM model also apply for the $\omega$CDM case,
showing that the Pad\'e approximants give a better description of the exact luminosity distance over the full interval considered, as one can
see in Fig.~\ref{wCDMDL}.
Exactly as in the $\Lambda$CDM case, the best approximations are given by $P_{2 1}$, $P_{22}$ and $P_{3 2}$.
To conclude, Figs.~\ref{LCDMDL} and~\ref{wCDMDL} clearly show that, provided we are given data over a large interval of values for the cosmological redshift $z$,
it would be better to fit the observed luminosity distances with a rational function,
in order to get a more realistic function that fits such data over the whole interval.
By the same reasoning, the use of Pad\'e approximants seems to be also more convenient in order to infer the evolution of  $d_L$
from knowledge of the CS. In particular, it seems that the Pad\'e approximants $P_{2 1}$, $P_{2 2}$ and $P_{3 2}$ give the best approximations,
which strongly suggests that the order of the numerator and that of the denominator for these models should be very close to each other,
with the former possibly being greater than the latter.
Given this fact,  in the next section we will give a quantitative analysis  of different Pad\'e approximations for the luminosity distance,
by comparing them with the astronomical data. In this way we get the best values for the CS parameters by a direct fit using different forms.
As we will see, this novel approach can give better bounds on the parameters, and takes better account of more distant objects,
as the Pad\'e approximation over a large interval is more reliable than Taylor's technique.

The above considerations suggest  some theoretical conclusions to build up viable Pad\'e rational functions.
Here, we formalize a {possible} recipe to determine which Pad\'e rational functions are favored  {with respect to} others.
First, the Pad\'e function should smoothly evolve in the redshift range chosen for the particular cosmographic analysis.
Naively, this suggests that any possible Pad\'e approximant should not have  singularities in the observable redshift intervals.
Moreover, any Pad\'e {approximant for $d_{L}$} must be positive definite and cannot show negative regions,
otherwise the definition of magnitude would not hold at all.
 {Finally, we expect that the degree of the numerator and that of the denominator should be close, with the former a little greater than the latter.}
Keeping this in mind, we are ready to perform our experimental analyses.
To do so, we consider some Pad\'e expansions as reported in the following sections.

\section{Experimental analysis with Pad\'e functions}\label{Sect:DataSet}

In this section we present the main aspects of our experimental analysis. We illustrate how we
directly fit general expressions of $d_L$, in terms of different types of approximations, i.e. Taylor
(standard cosmographic approach), auxiliary variables and Pad\'e expansions (our novel cosmographic technique).
In general all Pad\'e approximations, due to their rational forms, may show spurious singularities for certain
values of the redshift $z$ lying in the interval of data. In other words, the need of constructing precise Pad\'e
approximations which are not plagued by divergences due to poles, is actually one of the tasks of our analysis.
In particular, a simple manner to completely avoid such a problem consists in the choice of suitable
priors for the free parameters, built up \emph{ad hoc}, shifting any possible poles to future time cosmological
evolution. We show that data are confined inside intervals of the form $z\in[0,\infty)$, whereas possible divergences
of Pad\'e functions are limited to future
times, i.e. $z\leq-1$, and hence do not influence our experimental analysis.
Moreover, the cosmological
priors adopted here are perfectly compatible with the ones proposed in several previous papers and do not influence the numerical outcomes.
This shows that the Pad\'e method does not reduce the accuracy in  fitting procedures  and it is a good candidate to improve standard methods of cosmographic analyses. Thus, let us investigate
the improvements of the Pad\'e treatments with respect to standard techniques.
To do so, we denote the cosmographic parameters
by a suitable vector ${\theta}$, whose dimension changes depending on how many coefficients we are going
to analyze in a single experimental test. Estimations of the cosmographic parameters have been performed
through Bayesian techniques and best fits have been obtained by maximizing the likelihood function, defined as
\begin{equation}\label{jhfdkjdf}
    \mathcal{L}
\propto \exp (-\chi^2/2 )\,,
\end{equation}
where $\chi^2$ is the common {\it (pseudo) $\chi$-squared function}, whose form is explicitly determined for
each data set employed. Maximizing the likelihood function leads to minimizing the pseudo-$\chi$-squared
function and it can be done by means of a direct comparison with each cosmological data set.

For our purposes, we describe three statistical data sets, characterized by different maximum order of parameters,
providing  a hierarchy among parameters. This procedure leads to a broadening of the sampled distributions
if the whole set of parameters is wider, i.e. if the dimension of ${\theta}$ is higher. As a consequence, the numerical
outcomes may show deeper errors, which may be healed by means of the above cited priors. We make use of
the supernova union 2.1 compilation from the  \emph{Supernovae Cosmology Project}~\cite{Suzuki:2011hu}, i.e.
free available data of the most recent and complete supernova survey. Further, we employ a Gaussian prior
on the present-time Hubble parameter, i.e. $H_0 = 73.8 \pm 2.4 \,\text{km/s/Mpc}$~\cite{Riess:2011yx} implied
by the \emph{Hubble Space Telescope} (HST) measurements, and we also consider the almost model independent Baryonic Acoustic Oscillation, {\it BAO ratio},
as proposed in~\cite{Percival:2009xn}. In addition, we use relevant measurements of the Hubble parameter $H(z)$
at 26 different redshifts spanning from $z=0.09$ to
$z=2.3$~\cite{Simon:2004tf,Stern:2009ep,Moresco:2012jh,Gaztanaga:2008xz, Scrimgeour:2012wt, Busca:2012bu},
commonly named \emph{Observational Hubble Data} (OHD) or differential Hubble measurements.

The cosmological priors that we have employed here are summarized in Tab.~\ref{tab:priors}, in which we report the largest numerical interval developed for any single variable.

\begin{widetext}
\begin{center}
\newlength{\mywidth}
\setlength{\mywidth}{0.54\textwidth}
\begin{table}
\begin{center}
\begin{tabular}{ccccc}
\begin{tabular*}{\mywidth}{cccccc}
\hline\hline
\begin{tabular}{rclclrcl}
\phantom{ccccc} & Flat priors & \phantom{cc} & \vline & \phantom{cc} & Additional  & Constraints &  \\  \hline 
$0.5\quad <$      &       $h$     & $< \quad 0.9$ & \vline & \phantom{cc} & \phantom{cc} & & \\
$0.001 \quad < $&$\Omega_{\rm b}h^2$ & $< \quad 0.09$ & \vline & \phantom{cc}  &  $\Omega_{k}\quad =$&$0$ & \phantom{cc}\\
$0.01 \quad <$&$\Omega_{\rm dm}h^2$ & $< \quad 0.25$ & \vline & \phantom{cc}  &  $\Omega_{m}\quad<$&$0.5$ & \phantom{cc}  \\
$-1000\, \quad <$&$ {\tilde \Theta}$ & $< \quad 1000\,$ & \vline &\phantom{cc} & \phantom{cc}& &\\
\end{tabular}\\\hline
\end{tabular*}
\end{tabular}
\end{center}
\caption{Priors imposed on the free parameters involved in the Bayesian analysis for all cosmographic tests here employed. The parameter $h$ is the normalized Hubble rate, while $\tilde\Theta$ indicates a generic cosmographic coefficient ($q_0,j_0,s_0,\ldots$). We also report geometrical consequences on scalar curvature and the whole matter density.}\label{tab:priors}
\end{table}
\end{center}
\end{widetext}

Now, we are ready to investigate whether and how much Pad\'e approximants are favored for estimating bounds on the late time universe.
To better illustrate the procedure, we report below the $\chi^2$ function for each of the data sets adopted in the numerical analysis.

\subsection{Supernova type Ia compilation}

Type Ia supernovae observations have been extensively analyzed during the last decades for parameter-fitting
of cosmological models. They are considered as standard candles, i.e. quantities whose luminosity curves are intimately related
to distances. In our work, we employ the most recent survey of supernovae compilations, namely union 2.1~\cite{Suzuki:2011hu},
which extends previous versions union and union 2 data sets~\cite{Amanullah:2010vv,Kowalski:2008ez}. Here, systematics is reduced
and does not influence numerical outcomes, as for previous surveys.

The standard fitting procedure relies on using a Gaussian {\it $\chi$-squared} function, evaluating differences between
theoretical and observational distance modulus $\mu(z_i; {\theta}) - \mu_{obs}(z_i)$.
Nevertheless, the presence of nuisance parameters as the Hubble factor $ H_0$ and absolute magnitude $M$ enforces
to marginalize over. Straightforward calculations provide~\cite{GoliathAmanullah}
\begin{equation}
\chi^2_{SN} = A - \frac{B^2}{C} + \log \left( \frac{C}{2\pi}\right)\,,
\end{equation}
where we defined
\begin{eqnarray}
 A &=& {\bf x}^T \mathcal{C}^{-1} {\bf x}\,, \nonumber \\
 B &=& \sum_i (\mathcal{C}^{-1}{\bf x})_i\,,  \\
 C &=& \text{Tr}[\,\mathcal{C}^{-1}\,]\,.  \nonumber
\end{eqnarray}
Here $\mathcal{C}$ represents the covariance matrix of observational data, including statistical and systematic errors
as well, and the $i$-th component of the vector ${\bf x}$ given by
\begin{equation}
 {\bf x}_i= 5\log_{10}\left(\frac{d_L(z_i; { \theta})}{\text{Mpc}}\right) + 25 - \mu_{obs}(z_i)\,.
\end{equation}

\subsection{Ratio of baryonic acoustic oscillation}\label{subsec:BAO}

The ratio of baryonic acoustic oscillation (BAO) is slightly model dependent~\cite{Durrer2011}, since acoustic scales actually depend
on the redshift (drag time redshift), inferred from a first order perturbation theory. However, baryonic acoustic oscillations
determined from~\cite{Percival:2009xn} have been found in terms of a model independent quantity, i.e.
\begin{equation}
 \mathcal{B_R}\equiv D_V(0.35)/ D_V(0.20) = 1.736 \pm 0.065\,,
\end{equation}
where the volumetric distance is defined as
\begin{equation}\label{defDV}
 D_V(z) = \left[c \,z\, d_L(z)^2/(1+z)^2H(z) \right]^{1/3}\,.
\end{equation}
The BAO ratio $\chi$-squared function is simply given by
\begin{equation}
 \chi^2_{BAOr}(\theta) = \frac{(D_V(0.35; \theta)/ D_V(0.20; \theta) - 1.736)^2}{0.065^2}\,.
\end{equation}

\noindent We describe below the procedure to compute $D_{V}$ by means of the Pad\'e expansions for $d_{L}$. First, one needs to compute the approximation of $H(z)$ in terms of $d_L$ by inverting Eq. (\ref{defDL}). It follows
\begin{equation}\label{HfromDL}
H(z)=\left[\frac{\rm d}{{\rm d} z}\left(\frac{d_L}{1+z}\right)\right]^{-1}\,.
\end{equation}
Afterwards, inserting $H(z)$ from Eq. ($\ref{HfromDL}$) and the Pad\'e expressions for $d_{L}$ into Eq. (\ref{defDV}), one obtains the corresponding approximations for $D_V$, as reported in Appendix A.

\subsection{Direct Hubble measurements}\label{ohd}

We use 26 independent OHD data from~\cite{Simon:2004tf,Stern:2009ep,Moresco:2012jh,Gaztanaga:2008xz, Scrimgeour:2012wt, Busca:2012bu}, as reported in the appendix of this work.
We use those data, following~\cite{Simon:2004tf,Stern:2009ep,Moresco:2012jh}, in which a novel approach to track the universe expansion history has
been proposed, employing massive early type galaxies as cosmic chronometers~\cite{JimenezLoeb}. The technique
allows one to estimate the quantity $dt/dz$, sometimes referred to as \emph{differential time}, which is related to the Hubble rate by
\begin{equation}\label{njvnjfk}
H(z) = - (1+z)^{-1}dz/dt\,.
\end{equation}
Bearing in mind Eq.~(\ref{njvnjfk}), a preliminary list of 19 numerical outcomes has been found, whereas the
other 7 data have been determined from the study of galaxy surveys: 2 from~\cite{Gaztanaga:2008xz}, 4 from the wiggle Z collaboration~\cite{Scrimgeour:2012wt}
and one more from~\cite{Busca:2012bu}. All Hubble estimates are uncorrelated, therefore
the $\chi$-$squared$ function is simply given by
\begin{equation}
 \chi^2_{OHD}({\theta})  = \sum_{i} \frac{(H(z_i; {\theta}) - H_{obs}(z_i))^2 }{\sigma_i^2}\,.
\end{equation}
In the Appendix~\ref{appA}, as already stressed, we provide Tab. \ref{table:OHD}, where we summarize the OHD data used in this paper.

\subsection{The fitting procedure}\label{fitting}

Due to the fact that the different data sets are uncorrelated, the total $\chi$-$squared$ function is given by
\begin{equation}
 \chi^2({\theta}) = \chi^2_{\text{SN}} + \chi^2_{\text{OHD}} + \chi^2_{\text{BAOr}} + \chi^2_{\text{HST}}\,.
\end{equation}
The best fit to the data is given by those parameters that maximize the likelihood function $\mathcal{L} \propto \exp (-\chi^2/2)$. We obtain them and their
respective confidence intervals
by using a Metropolis-Hasting Markov Chain Monte Carlo (MCMC) algorithm~\cite{Hastings70,Metropolis53} with the publicly
available CosmoMC code~\cite{Lewis:2002ah,cosmomc_notes}.
We run several independent chains and to probe their
convergence we use the Gelman-Rubin criteria $R \sim \text{``mean of chains variances''} / \text{``variance of chains means''}$ \cite{Gelman:1992zz}
with $R-1 < 0.01$. We accurately modify the priors for each ${\theta}$, within the interval of values reported in Tab.~\ref{tab:priors}.

\section{Estimation of the cosmographic series} \label{ParameterEstimations}

\begin{figure}
\begin{center}
\includegraphics[width=3.4in]{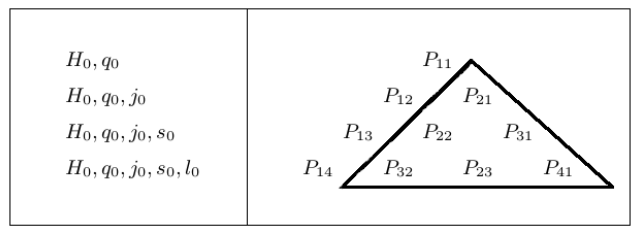}
{\small \caption{The Pad\'e approximants used for the different ${\theta}$. The approximations enclosed in the triangle give
conclusive results.}
\label{fig:PA}}
\end{center}
\end{figure}

\begin{figure*} [ht]
\begin{center}
\includegraphics[width=2in]{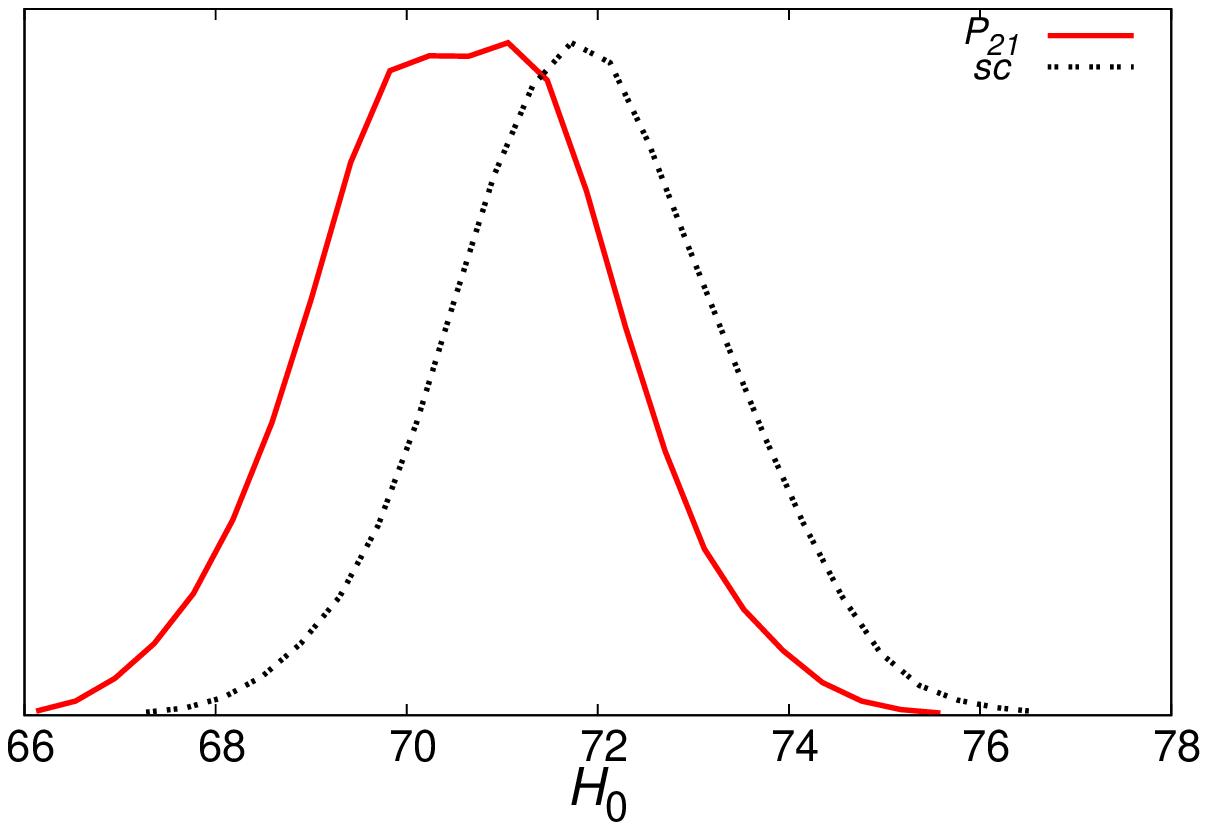}
\includegraphics[width=2in]{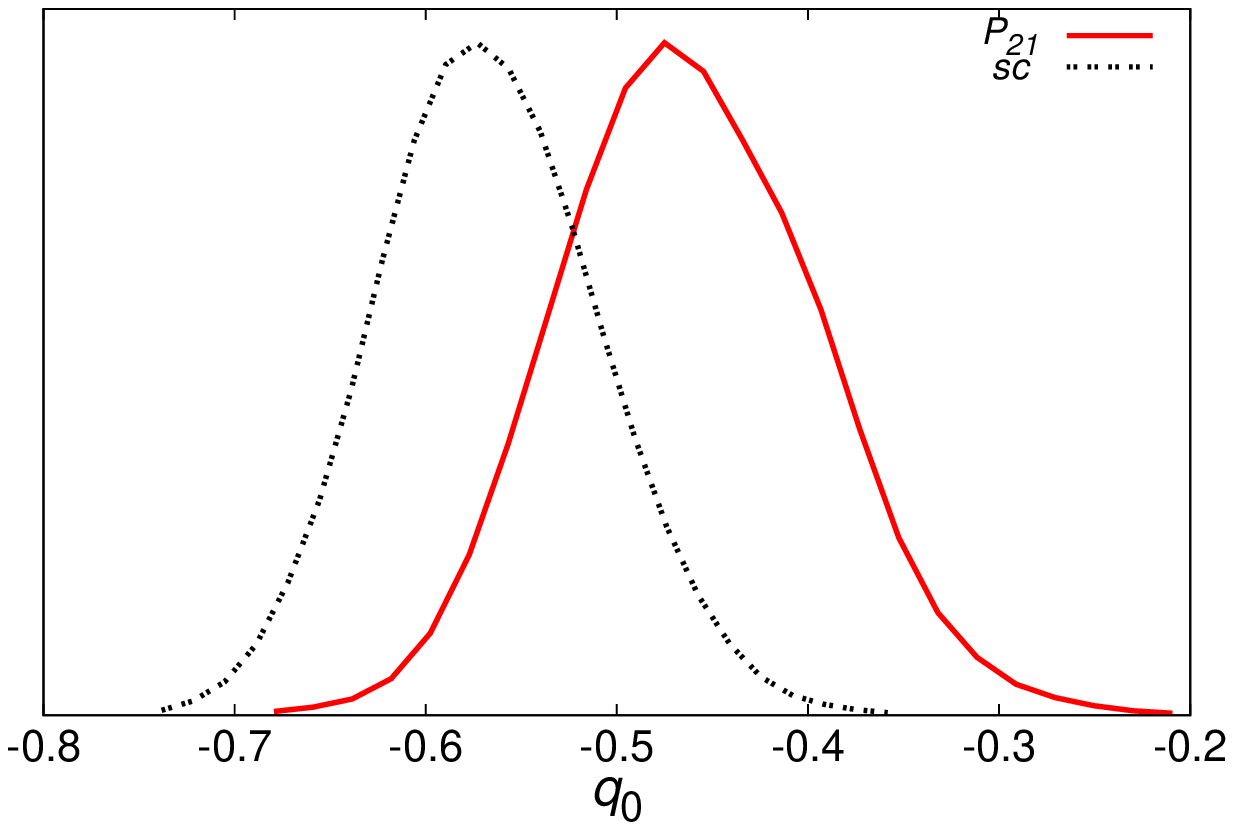}
\includegraphics[width=2in]{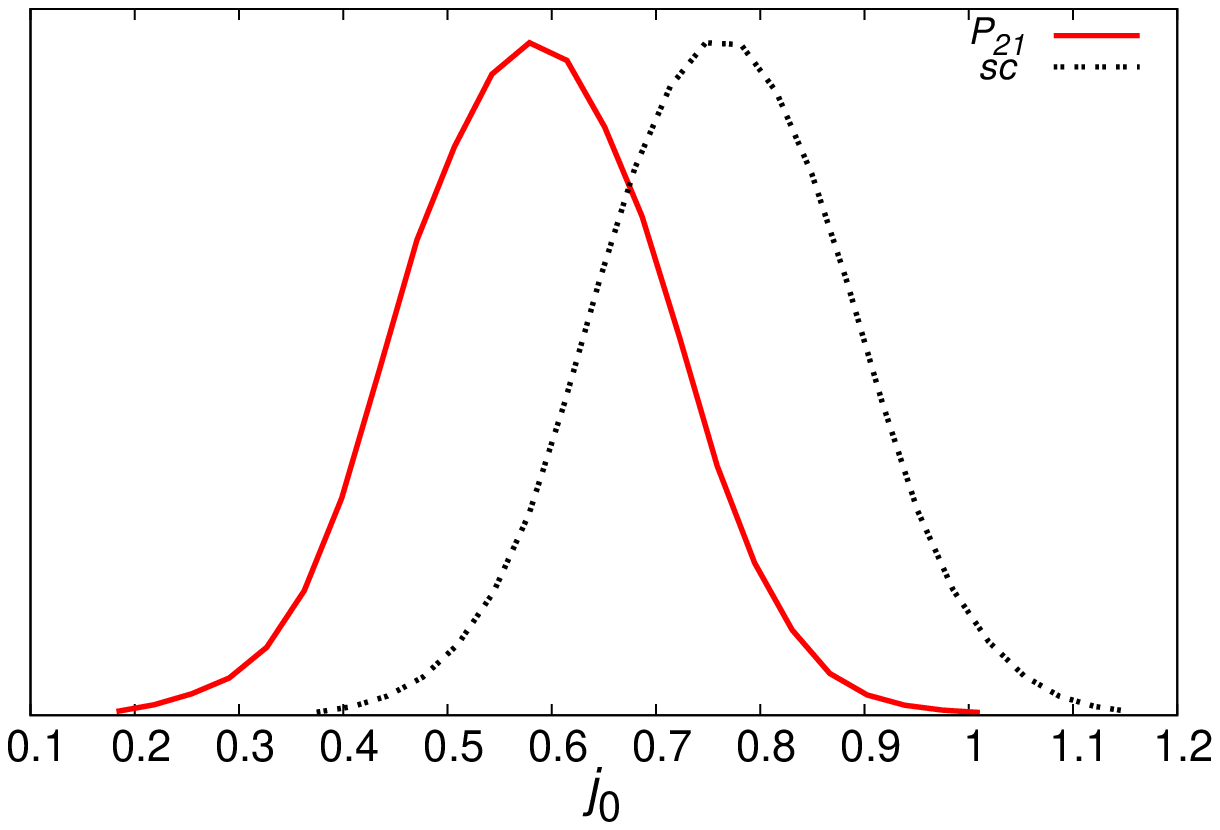}
{\small \caption{(color online) 1-dimensional marginalized posteriors for the
parameter set $\mathcal{A}$. $SC$ stands for standard cosmography.}
\label{fig:1dimModelA}}
\end{center}
\end{figure*}

For the parameters estimation  we will use the CS combined
in three sets with different maximum order of
parameters:

\begin{eqnarray}
 \mathcal{A} &=& \{ H_0, q_0, j_0\}\,, \nonumber\\
 \mathcal{B} &=& \{ H_0, q_0, j_0, s_0 \}\,,\\
 \mathcal{C} &=& \{ H_0, q_0, j_0, s_0, l_0 \}\,.  \nonumber
\end{eqnarray}

For the parameters set $\mathcal{A}$ the corresponding Pad\'e approximants are $P_{12}$ and $P_{21}$,
as shown in Fig.~\ref{fig:PA}. Of those approximants, only $P_{21}$ gives
conclusive results. For $\mathcal{B}$ we obtain conclusive results for $P_{31}$ and $P_{22}$
and for $\mathcal{C}$ for $P_{32}$, $P_{23}$ and $P_{41}$.

In Tables~\ref{table:1DResults1},~\ref{table:1DResults2}, and~\ref{table:1DResults3} we show
the best fits and their $1\sigma$-likelihoods for the  parameters sets
$\mathcal{A}$, $\mathcal{B}$ and $\mathcal{C}$ respectively.
We also show the estimated CS obtained by the standard cosmography ($SC$) or Taylor approach.
We worked out the $\Lambda$CDM model, which is for our purposes and the redshifts involved sufficiently described by two parameters: $\Omega_m h^2$ and $\Theta$.
Here $\Theta$ is defined as 100 times the ratio of the sound horizon to the angular diameter distance at recombination,
while as usual $\Omega_m$ is the abundance of matter density (both baryonic $\Omega_b$ and dark matter $\Omega_{dm}$), and $h$ is
the dimensionless Hubble parameter, as reported in Tab.~\ref{tab:priors}.

The best fits, using the same data sets as above, are given by $\Omega_m h^2 = 0.148${\tiny ${}_{-0.010}^{+0.012}$} and
$\Theta = 1.041${\tiny ${}_{-0.010}^{+0.011}$}. From these values and the formulas
\begin{eqnarray} \label{LCDMCS}
q_0 &=&  -1 + \frac{3}{2}\Omega_m\,, \nonumber\\
j_0  &=&  1\,, \nonumber\\
s_0 &=&  1 - \frac{9}{2} \Omega_m\,, \\
l_0 &=& 1 +  3 \Omega_m - \frac{27}{2}\Omega_m^2\,, \nonumber
\end{eqnarray}
which are valid only for flat $\Lambda$CDM model, we have also estimated the cosmographic parameters and
we report them in Tables~\ref{table:1DResults1},~\ref{table:1DResults2}, and~\ref{table:1DResults3}.

\begin{table*}
\caption{{\small Table of best fits and their likelihoods (1$\sigma$) for the parameter set $\mathcal{A}$. $SC$ stands for the standard cosmography approach and
the $\Lambda$CDM derived columns are the parameters inferred assuming as valid the $\Lambda$CDM model.}}

\begin{tabular}{c|c|c|c}

\hline\hline

{\small $\quad$ Parameter $\quad$}  &   {\small $\qquad$ $P_{21}$ $\qquad$ }
                                    &   {\small  $\qquad$ $SC$ $\qquad$}         & {\small $\quad$ $\Lambda$CDM derived $\quad$}\\

\hline

{\small$H_0$}       & {\small $70.64$}{\tiny ${}_{-2.63}^{+2.77}$}
                    & {\small $71.98$}{\tiny ${}_{-2.55}^{+2.48}$}        & {\small $71.68$}{\tiny${}_{-2.16}^{+2.25}$}\\[0.8ex]

{\small$q_0$}       & {\small $-0.4712$}{\tiny ${}_{-0.1106}^{+0.1224}$}
                    & {\small $-0.5701$}{\tiny ${}_{-0.0928}^{+0.1057}$}      & {\small $-0.6117$}{\tiny${}_{-0.0365}^{+0.0401}$}\\[0.8ex]

{\small$j_0$}       & {\small $0.593$}{\tiny ${}_{-0.210}^{+0.216}$}
                    & {\small $0.766$}{\tiny ${}_{-0.207}^{+0.211}$}         & {\small $1$}\\[0.8ex]

\hline \hline

\end{tabular}

{\tiny Notes.
a. $H_0$ is given in Km/s/Mpc units.
}

\label{table:1DResults1}
\end{table*}

\begin{table*}
\caption{{\small Table of best fits and their likelihoods (1$\sigma$) for the parameter set $\mathcal{B}$. $SC$ stands for the standard cosmography approach and
the $\Lambda$CDM derived columns are the parameters inferred assuming as valid the $\Lambda$CDM model.}}

\begin{tabular}{c|c|c|c|c}

\hline\hline

{\small $\quad$ Parameter $\quad$}  &   {\small $\qquad$ $P_{31}$ $\qquad$ }  & {\small $\qquad$ $P_{22}$ $\qquad$ }
                                    &   {\small $\qquad$ $SC$ $\qquad$}         & {\small $\quad$ $\Lambda$CDM derived $\quad$}\\

\hline

{\small$H_0$}       & {\small $71.76$}{\tiny${}_{-3.46}^{+ 3.38}$}          & {\small $71.71$}{\tiny${}_{-3.15 }^{+3.35}$}
                    & {\small $72.53$}{\tiny ${}_{-3.51}^{+3.53}$}        & {\small $71.68$}{\tiny${}_{-2.16}^{+2.25}$}\\[0.8ex]

{\small$q_0$}       & {\small $-0.6483$}{\tiny${}_{-0.1623}^{+0.2589}$}      & {\small $-0.6767$}{\tiny${}_{-0.2580}^{+0.2395}$}
                    & {\small $-0.6642$}{\tiny${}_{-0.1963}^{+0.2050}$}      & {\small $-0.6117$}{\tiny${}_{-0.0365}^{+0.0401}$}\\[0.8ex]

{\small$j_0$}       & {\small $1.313$}{\tiny${}_{-0.917}^{+0.521}$}          & {\small $1.500$}{\tiny${}_{-1.009}^{+0.973}$}
                    & {\small $1.223$}{\tiny ${}_{-0.664}^{+0.644}$}          & {\small $1$}\\[0.8ex]

{\small$s_0$}       & {\small $0.425$}{\tiny${}_{-0.841}^{+1.079}$}          & {\small $0.681$}{\tiny${}_{-1.055}^{+2.367}$}
                    & {\small $0.394$}{\tiny ${}_{-0.731}^{+1.335}$}         & {\small $-0.165$}{\tiny${}_{-0.120}^{+0.109}$}\\[0.8ex]

\hline \hline

\end{tabular}

{\tiny Notes.
a. $H_0$ is given in Km/s/Mpc units.
}

\label{table:1DResults2}
\end{table*}

\begin{table*}
\caption{{\small Table of best fits and their likelihoods (1$\sigma$) for the parameter set $\mathcal{C}$. $SC$ stands for the standard cosmography approach and
the $\Lambda$CDM derived columns are the parameters inferred assuming as valid the $\Lambda$CDM model.}}

\begin{tabular}{c|c|c|c|c|c}

\hline\hline

{\small $\quad$ Parameter $\quad$}  &   {\small $\qquad$ $P_{41}$ $\qquad$ }  & {\small $\qquad$ $P_{32}$ $\qquad$ }
				     &   {\small $\qquad$ $P_{23}$ $\qquad$ }
                                    &   {\small $\qquad$ $SC$ $\qquad$}         & {\small $\quad$ $\Lambda$CDM derived $\quad$}\\

\hline

{\small$H_0$}       & {\small $71.56$}{\tiny${}_{-3.95}^{+3.95}$}          & {\small $71.83$}{\tiny ${}_{-3.64}^{+3.53}$}
		     & {\small $70.75$}{\tiny ${}_{-3.12}^{+3.41}$}
                    & {\small $71.38$}{\tiny ${}_{-3.68}^{+4.01}$}        & {\small $71.68$}{\tiny${}_{-2.25}^{+2.16}$}\\[0.8ex]

{\small$q_0$}       & {\small $-0.5516$}{\tiny${}_{-0.4556}^{+0.3190}$}      & {\small $-0.7189$}{\tiny ${}_{-0.4631}^{+0.3397}$}
                    & {\small $-0.5539$}{\tiny ${}_{-0.2171}^{+0.2966}$}
                    & {\small $-0.6173$}{\tiny ${}_{-0.3139}^{+0.3658}$}          & {\small $-0.6117$}{\tiny${}_{-0.0401}^{+0.0365}$}\\[0.8ex]

{\small$j_0$}       & {\small $0.721$}{\tiny${}_{-1.982}^{+2.489}$}          & {\small $1.959$}{\tiny ${}_{-2.516}^{+3.290}$}
                    & {\small $0.710$}{\tiny ${}_{-1.499}^{+1.389}$}
                    & {\small $0.949$}{\tiny ${}_{-1.686}^{+1.374}$}          & {\small $1$}\\[0.8ex]

{\small$s_0$}       & {\small $-1.060$}{\tiny${}_{-3.193}^ {+7.341}$}          & {\small $1.950$}{\tiny ${}_{-5.524}^{+14.072}$}
                    & {\small $-1.203$}{\tiny ${}_{-2.864}^{+3.073}$}
                    & {\small $-0.797$}{\tiny ${}_{-3.585}^{+2.962}$}          & {\small $-0.165$}{\tiny${}_{-0.109}^{+0.120}$}\\[0.8ex]

{\small$l_0$}       & {\small $4.43$}{\tiny${}_{-2.98}^{+19.79}$}          & {\small $8.14$}{\tiny ${}_{-7.05}^{+71.96}$}
                    & {\small $4.82$}{\tiny ${}_{-3.98}^{+13.07}$}
                    & {\small $4.47$}{\tiny ${}_{-3.76}^{+18.67}$}          & {\small $2.681$}{\tiny${}_{-0.277}^{+0.235}$}\\[0.8ex]

\hline \hline

\end{tabular}

{\tiny Notes.
a. $H_0$ is given in Km/s/Mpc units.
}

\label{table:1DResults3}
\end{table*}


\begin{figure*}
\begin{center}
\includegraphics[width=2in]{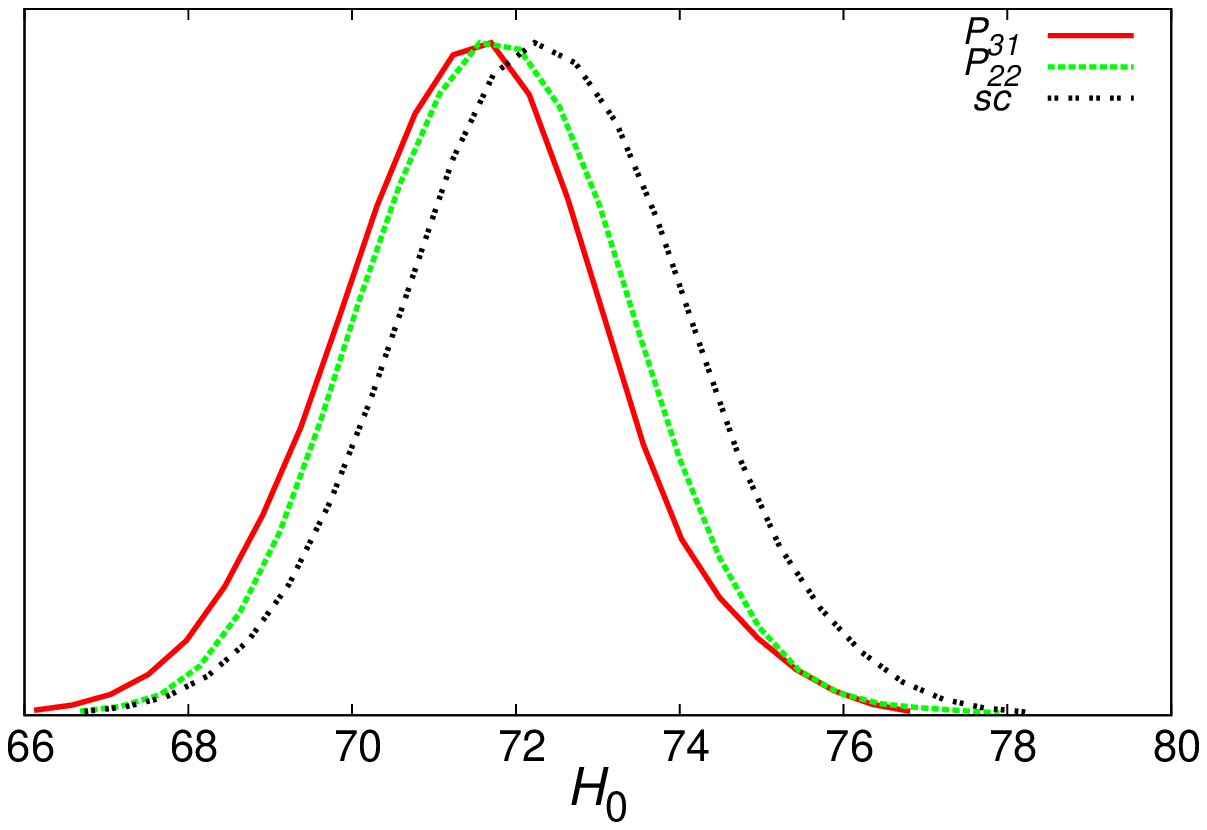}
\includegraphics[width=2in]{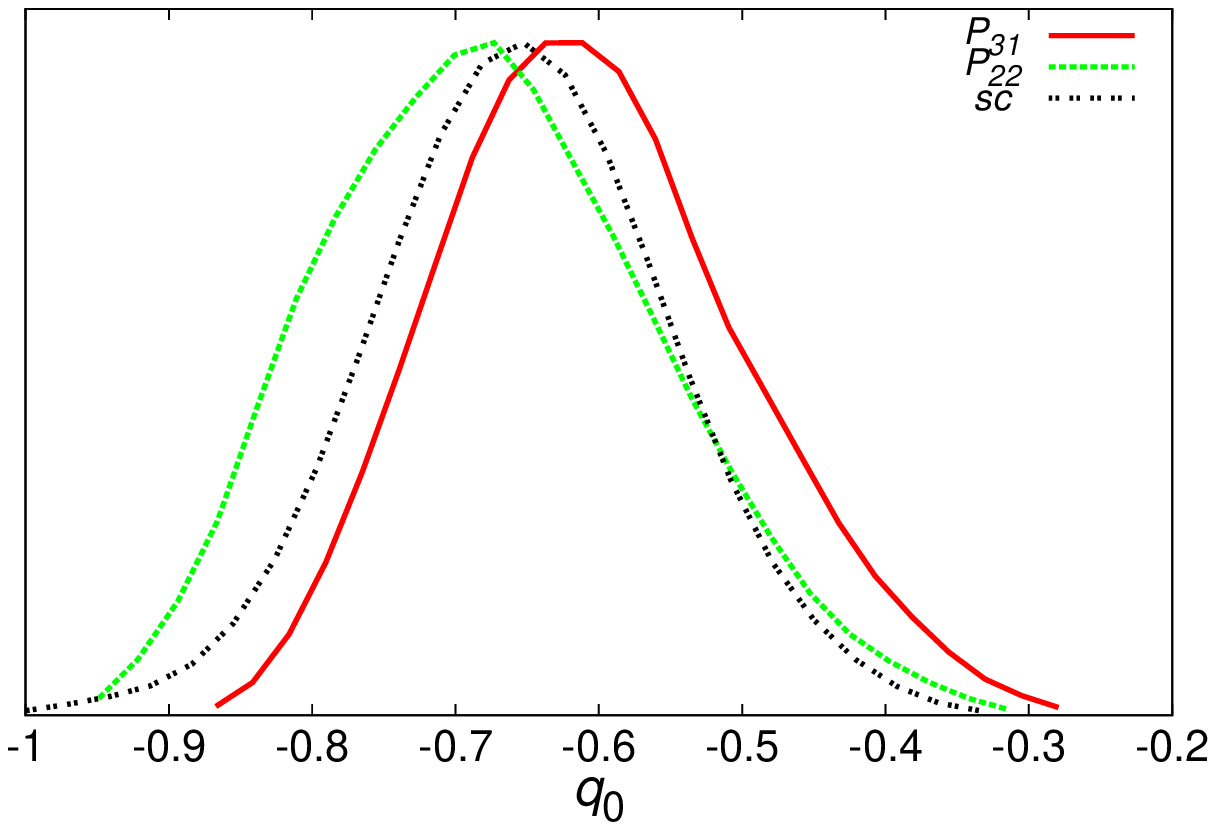}
\includegraphics[width=2in]{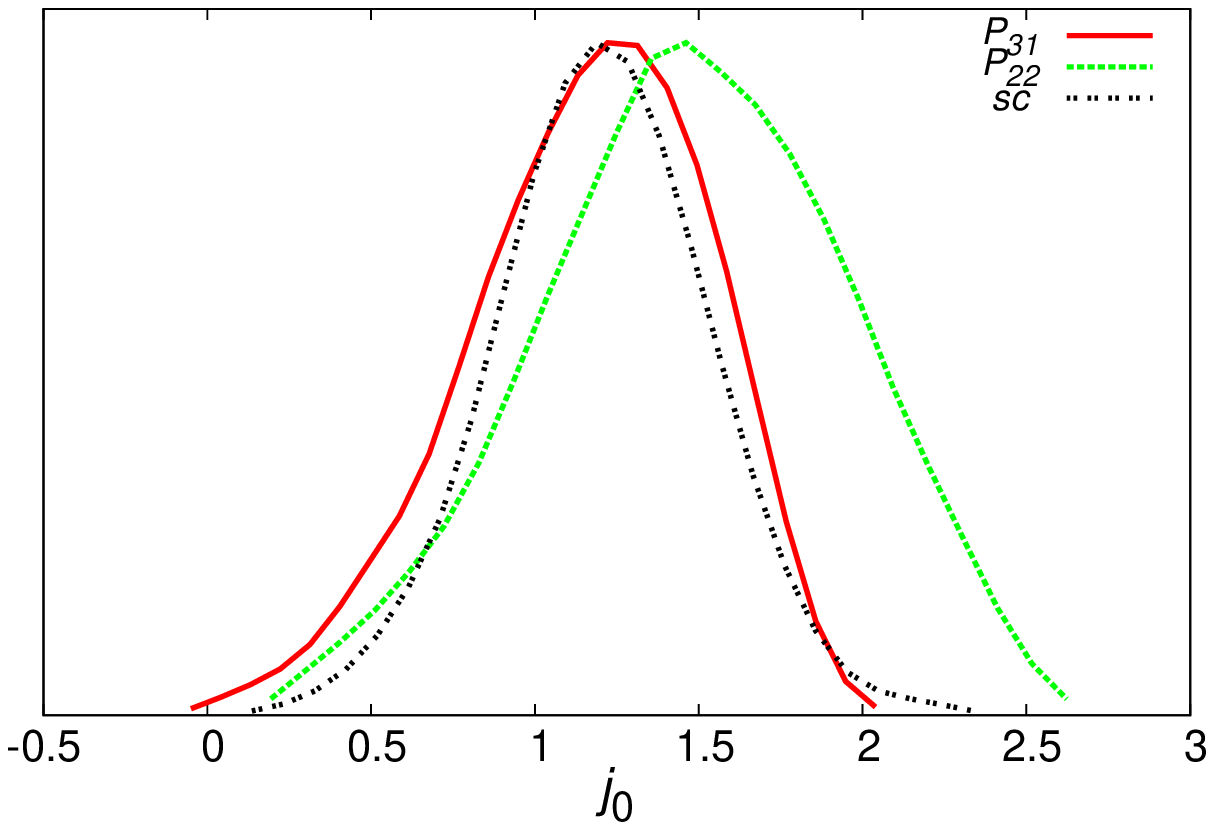}
\includegraphics[width=2in]{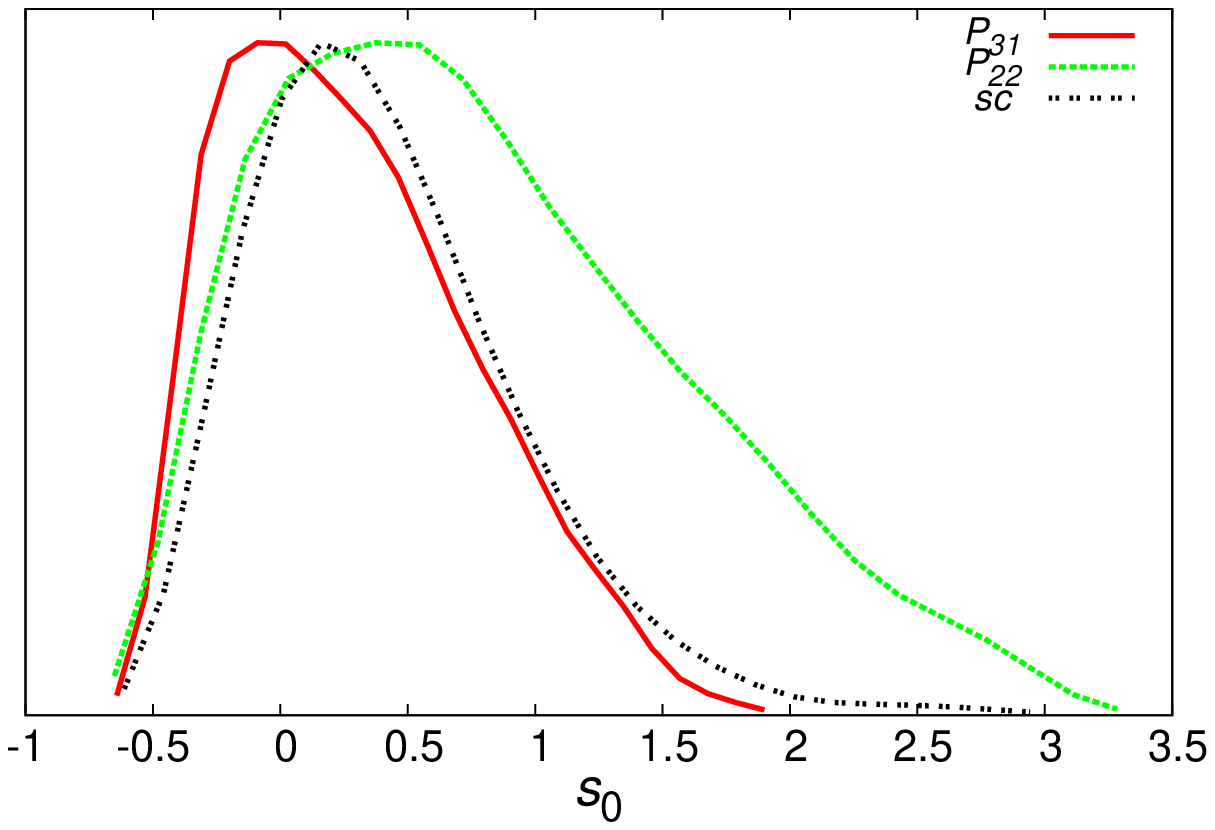}
{\small\caption{(color online) 1-dimensional marginalized posteriors for
parameter set $\mathcal{B}$. $SC$ stands for standard cosmography.}
\label{fig:1dim2ModelB}}
\end{center}
\end{figure*}

\begin{figure*}
\begin{center}
\includegraphics[width=2in]{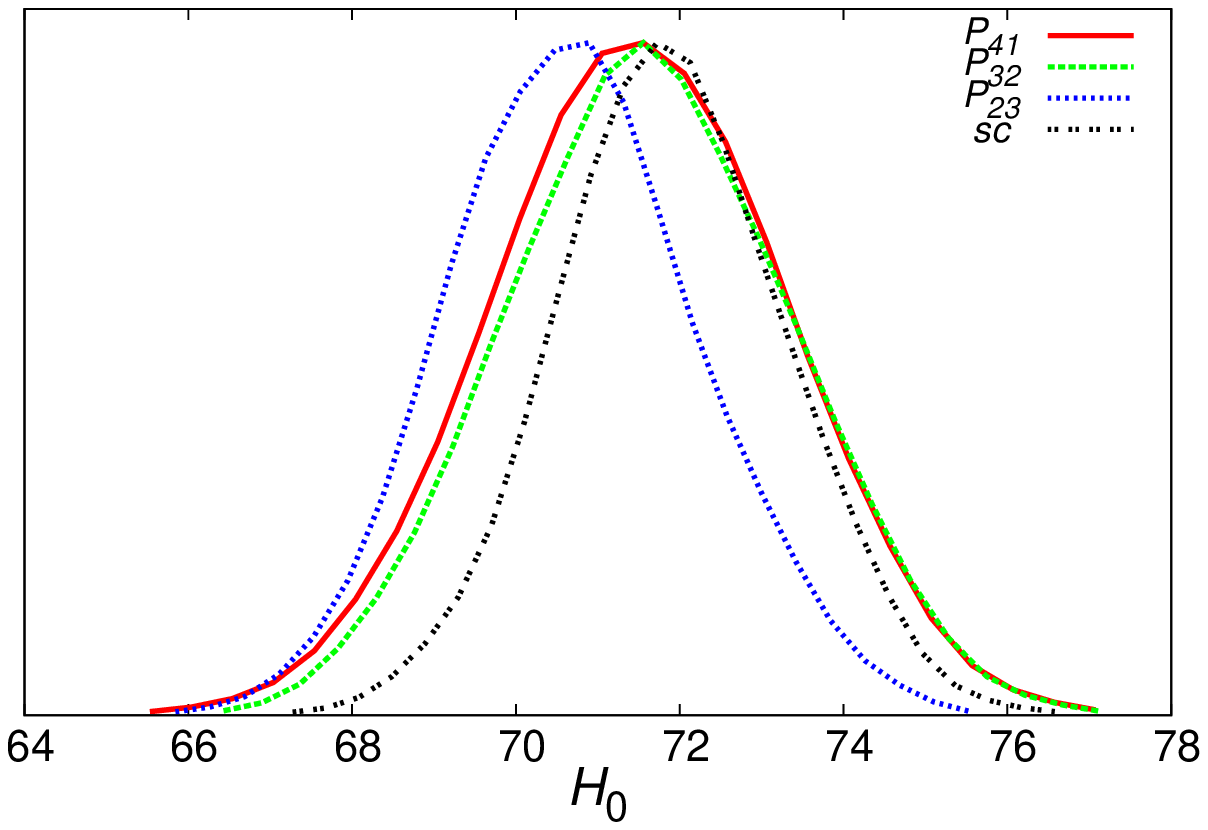}
\includegraphics[width=2in]{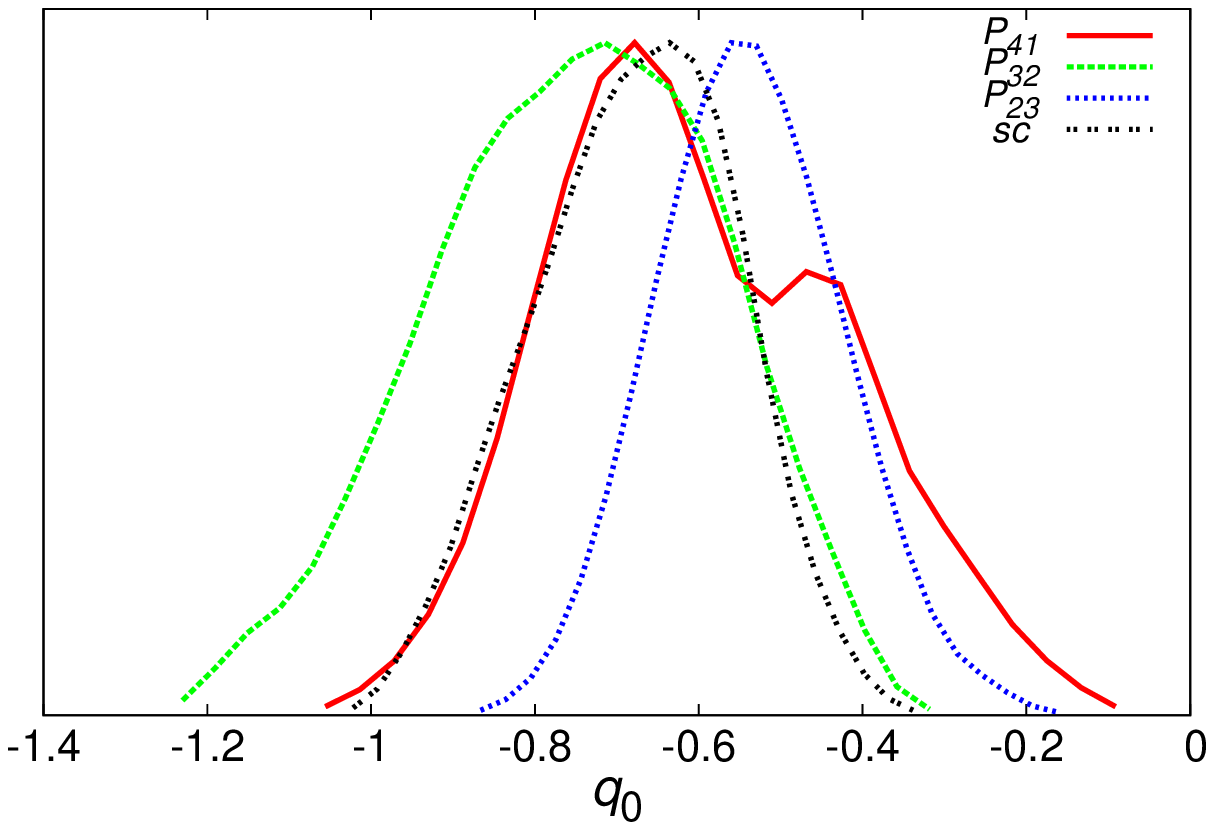}
\includegraphics[width=2in]{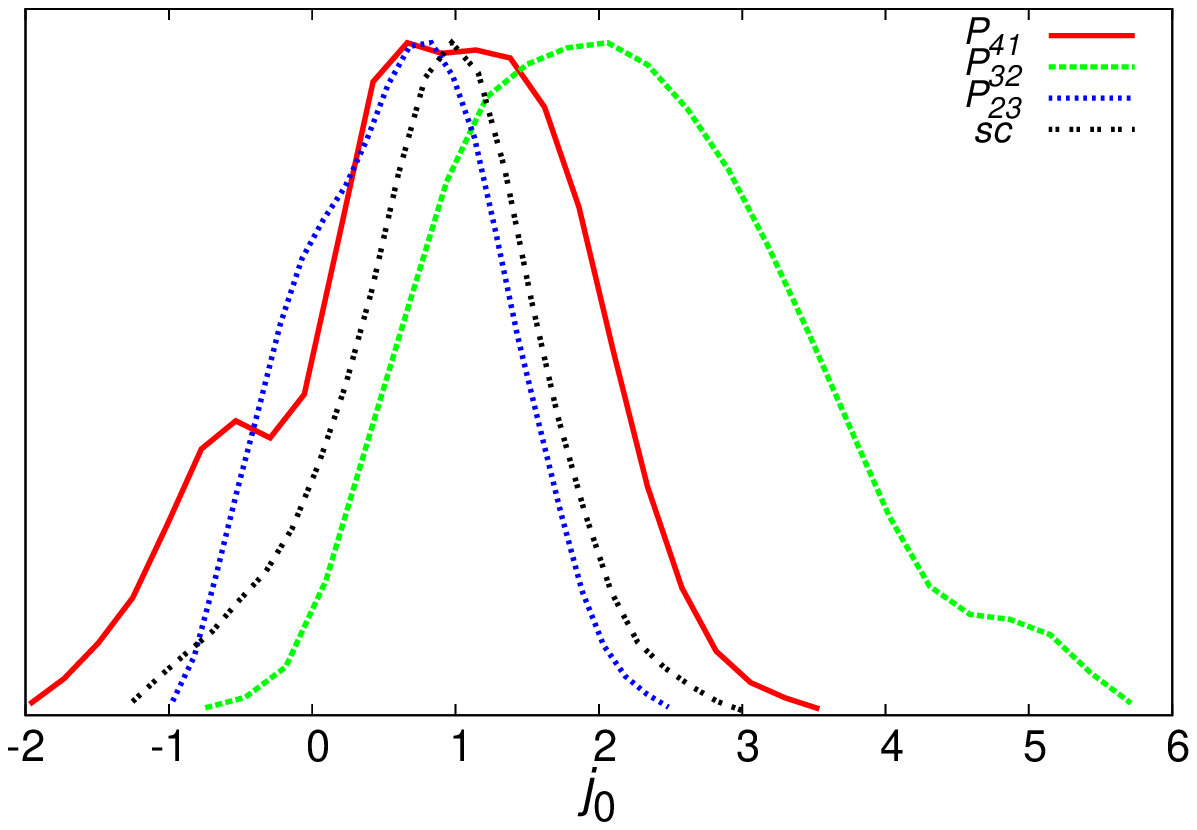}
\includegraphics[width=2in]{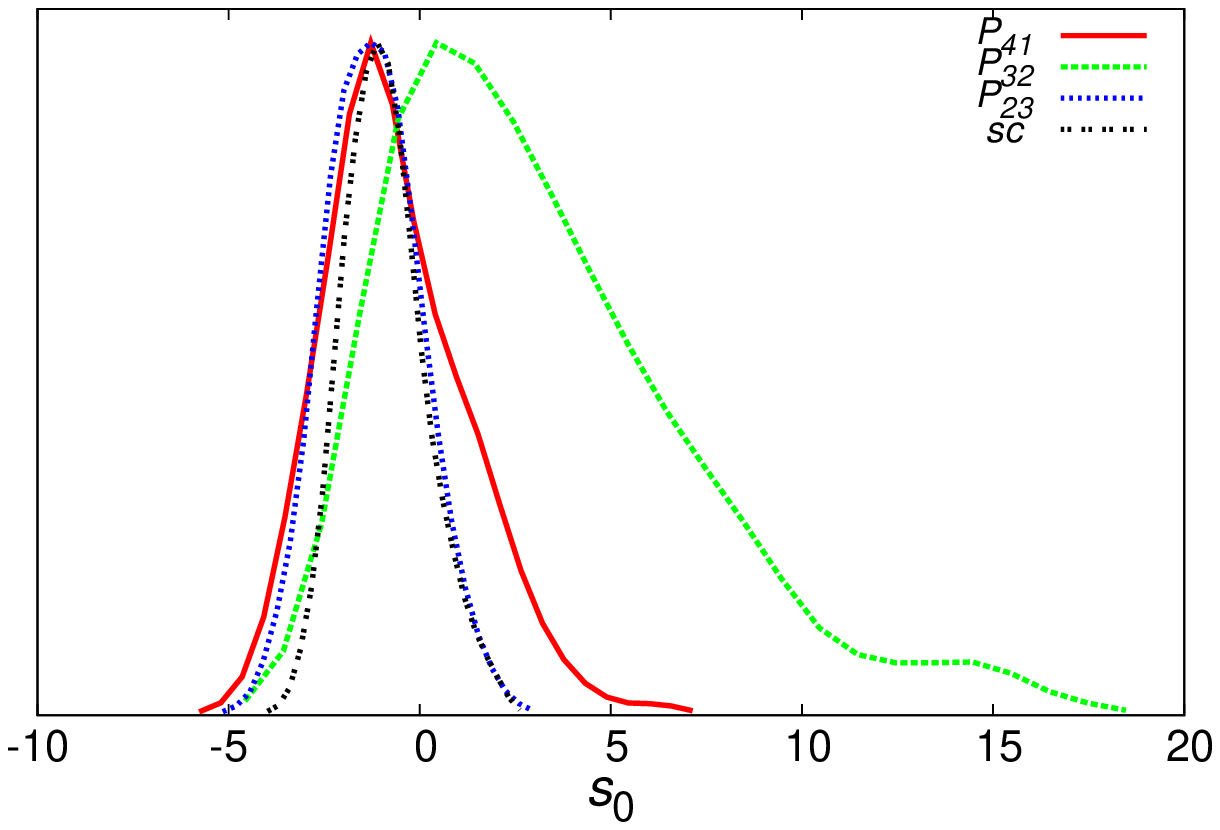}
\includegraphics[width=2in]{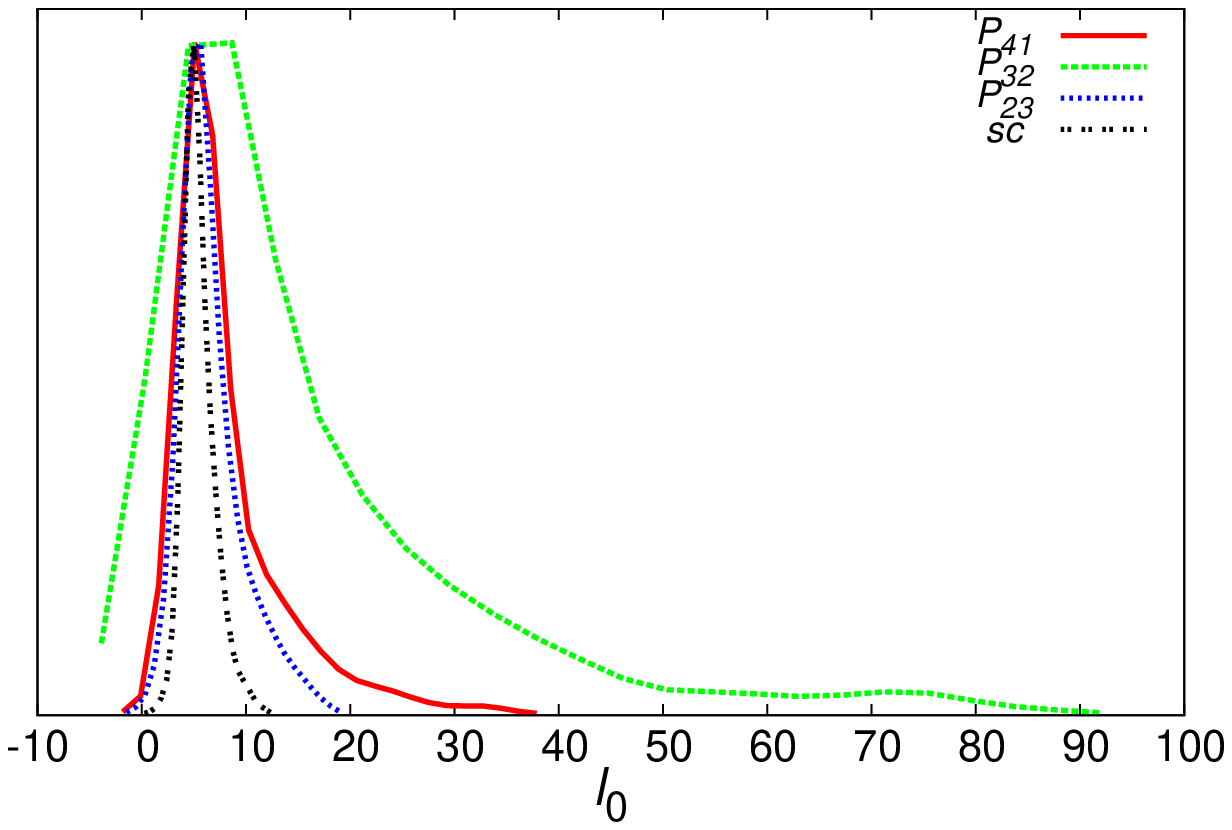}
{\small\caption{(color online) 1-dimensional marginalized posteriors for
parameter set $\mathcal{C}$. $SC$ stands for standard cosmography.}
\label{fig:1dim2ModelC}}
\end{center}
\end{figure*}

\begin{figure*}
\begin{center}
\includegraphics[width=3.4in]{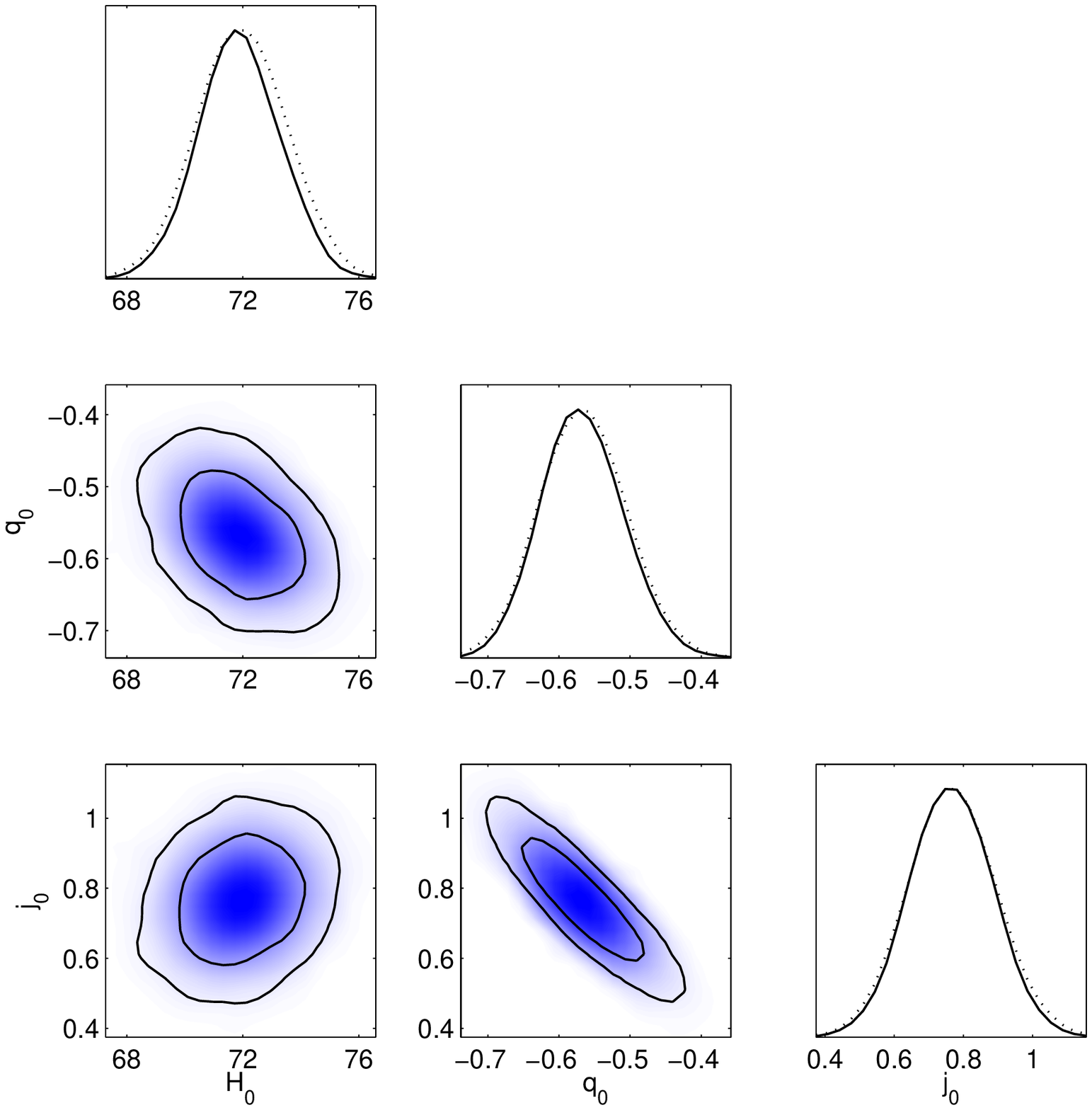}
{\small\caption{(color online) Marginalized posterior constraints for
parameter set $\mathcal{A}$ using $P_{21}$.}
\label{fig:Pade21}}
\end{center}
\end{figure*}

\begin{figure*}
\begin{center}
\includegraphics[width=3.4in]{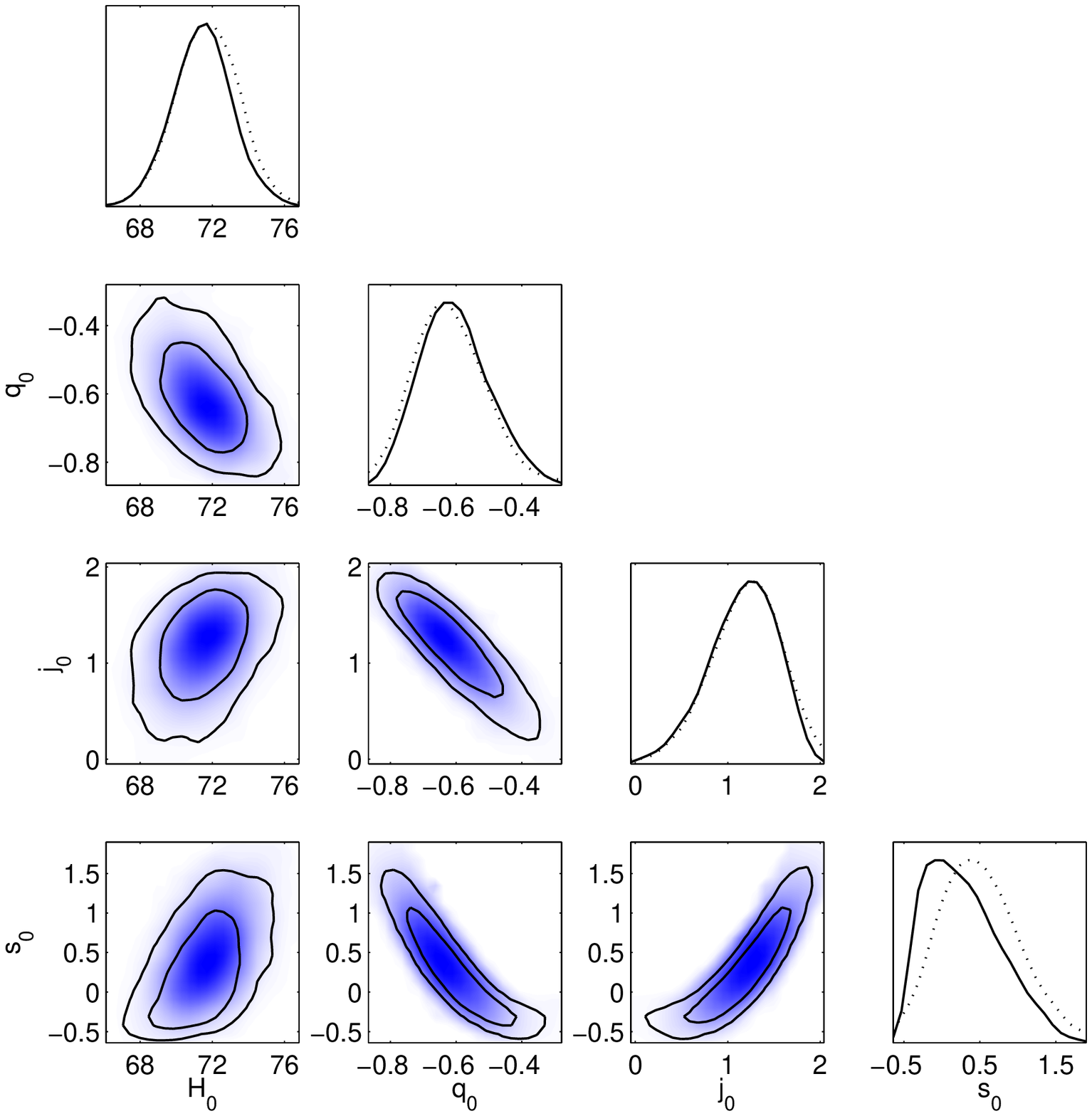}
{\small\caption{(color online) Marginalized posterior constraints for
parameter set $\mathcal{B}$ using $P_{31}$.}
\label{fig:Pade31}}
\end{center}
\end{figure*}

\begin{figure*}
\begin{center}
\includegraphics[width=3.6in]{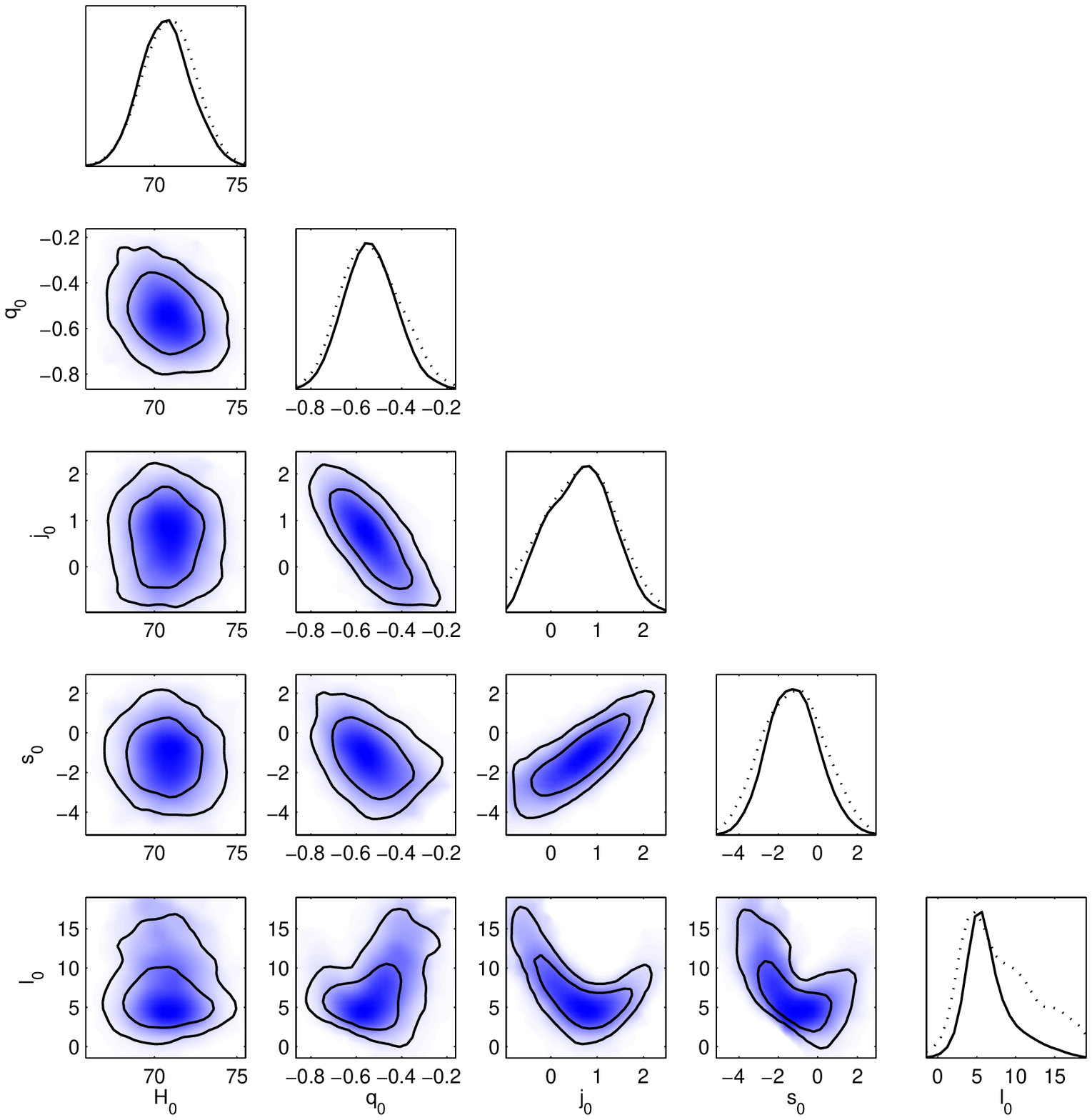}
{\small\caption{(color online) Marginalized posterior constraints for
parameter set $\mathcal{C}$ using $P_{23}$.}
\label{fig:Pade32}}
\end{center}
\end{figure*}

As we can observe from Figs.~\ref{fig:1dimModelA},~\ref{fig:1dim2ModelB} and~\ref{fig:1dim2ModelC},
the Pad\'e approximants give similar results to standard cosmography, with the advantage of the convergence properties discussed in the previous sections.
We note that in particular $P_{21}$, $P_{31}$ and $P_{23}$ draw better samples with narrower
dispersion. For this reason, we plot the contours for those approximants in Figs.~\ref{fig:Pade21},~\ref{fig:Pade31}
and~\ref{fig:Pade32}. It is remarkable that the same degeneracies
among the parameters are found in all cases, even in other cases  which have not been investigated here, see, e.g.~\cite{mine2}.

\section{Applications of Pad\'e's approach}\label{applicationsPade}

In flat FRW metric~(\ref{frw}), the
Friedmann equation for the energy density $\rho\equiv\sum_i\rho_i$ reads $3H^2 = 8\pi G \sum_i \rho_i$, where
the sum is over all cosmic species contributing to the whole energy budget. Afterwards, recovering Bianchi identities, one gets the continuity equation in the form
 $\dot{\rho_i} = -3H(1+w_i)\rho_i$, in the absence of energy transfer
among the different components. In order to determine a specific model, one must specify the
cosmic fluids and their equations of state~\cite{galaxyluongo}. Below we test some models by means of our cosmographic results, inferred from the Pad\'e formalism.

We deal with implicit propagation of errors, since it is convenient to work with expected values and variances of
the cosmographic parameters, instead of their probability distribution functions. Thus, for example
\begin{equation}
 \langle q_0 \rangle = \int q_0 p(q_0) dq_0\,,
\end{equation}
where $p(q_0) = f(q_0)/ \int f(q_0)dq$ and $f(q_0)$ is the non-normalized posterior distribution
found in section~\ref{ParameterEstimations} by the MCMC analysis.
The variance is
\begin{equation}
 \sigma^2_{q_0} =  \langle q_0^2 \rangle -  \langle q_0 \rangle^2\,,
\end{equation}
and similar equations hold for the other cosmological parameters.
For the Pad\'e approximants $P_{21}$, $P_{31}$  and  $P_{23}$ we obtain
\begin{itemize}

 \item Pad\'e approximant $P_{21}$:
\begin{eqnarray}
\langle q_0  \rangle&=& -0.4623 \pm 0.0677\,, \nonumber \\
\langle j_0 \rangle &=& 0.5834 \pm 0.1215\,, \label{P21expectedv}
\end{eqnarray}

 \item Pad\'e approximant $P_{31}$:
 \begin{eqnarray}
\langle q_0  \rangle&=& -0.6040 \pm 0.1051\,,  \nonumber \\
\langle j_0 \rangle &=& 1.1597 \pm 0.3690\,, \label{P31expectedv} \\
\langle s_0 \rangle &=& 0.2858 \pm 0.4866\,.  \nonumber
\end{eqnarray}

\item Pad\'e approximant $P_{23}$:
 \begin{eqnarray}
\langle q_0  \rangle&=& -0.7511 \pm 0.1737\,,  \nonumber \\
\langle j_0 \rangle &=& 2.1968 \pm 1.1828\,, \nonumber \\
\langle s_0 \rangle &=& 3.2038 \pm 4.0459\,,  \label{P32expectedv} \\
\langle l_0 \rangle &=& 15.9014 \pm 16.1370\,,  \nonumber
\end{eqnarray}

\end{itemize}
where the reported error values are the standard deviations of the probability distributions, $\sigma = \sqrt{\sigma^2}$.

Using these results we can approximate the probability distributions as Gaussians centered around their mean values
and with variance $\sigma^2$. Now we are ready to investigate the implications of the results obtained using Pad\'e on some relevant cosmological models.

\subsection{The case of the $\Lambda$CDM model}

Concerning the flat $\Lambda$CDM model, the parameters to estimate are only $H_0$ and $\Omega_m$.
It is easy to demonstrate that, while $H_0$ is actually one of the CS, the matter density can be related to $q_0$ as $\Omega_m(q_0) = 2 (q_0 +1) /3$.
We have found by the MCMC algorithm the distribution functions $f(q_0)$ for $q_0$, obtained using the results for the Pad\'e approximant $P_{21}$.
The expected value for $\Omega_m$ is given by
\begin{equation}
 \langle \Omega_m \rangle = \int \Omega_m(q_0) p(q_0) dq_0\,,
\end{equation}
and its variance is $ \sigma^2 =  \langle \Omega_m^2 \rangle -  \langle \Omega_m \rangle^2$, obtaining
\begin{equation} \label{LCDMOm0}
 \Omega_m = 0.36\pm 0.05\,.
\end{equation}
Using the results for the Pad\'e approximant $P_{31}$ we obtain, from $\Omega_m(q_0) = 2 (q_0 +1) /3$, see Eqs.~(\ref{LCDMCS}),
\begin{equation}
 \Omega_m(q_0) = 0.26\pm 0.07\,. \label{LCDMOm1}
\end{equation}
This procedure leads to a projection from a 4-parameter model (the Pad\'e $P_{31}$) to a 2-parameter model (late time flat-$\Lambda$CDM model), providing a broadening on the estimated parameters.
In analogy to the case in which $\Omega_m=\Omega(q_0)$, it is easy to show that $\Omega_m(s_0) = 2(1-s_0)/9$. Keeping in mind this expression, we obtain:
\begin{equation}
 \Omega_m(s_0) = 0.16\pm 0.11\,. \label{LCDMOm2}
\end{equation}
The combination of the two results~(\ref{LCDMOm1}) and~(\ref{LCDMOm2}) should give tighter constraints.
If the probability distribution functions of $s_0$ and $q_0$ are
independent, the distribution function of $\Omega_m$ is simply  the
product of the two distributions $\Omega_m(q_0)$ and  $\Omega_m(s_0)$.
If we further assume Gaussian distributions, all the statistical information
is given by Eqs.~(\ref{LCDMOm1}) and~(\ref{LCDMOm2}), obtaining a rough estimate:
\begin{equation}\label{mah}
 \Omega_m \simeq 0.23 \pm 0.06\,.
\end{equation}
In this way, we did a 3-parameter to 2-parameter projection. We cannot do anything better than Eq.~(\ref{mah}) for the flat-$\Lambda$CDM model due to the fact that $j_0$ is fixed to $j_0=1$.

As an example, to go beyond the case $j_0=1$, one can consider a generic additional cosmic component $X$, relevant at late times.
To do so, its equation of state parameter should lie in the interval
$-1 < w_X < 0$; but to avoid large degeneracies with a cosmological constant or with dust fluids we cannot be very close to $-1$ or to $0$.
A possible example is offered by the scalar curvature $\Omega_k$, that we neglected in all our previous numerical outcomes.
In such case, one can choose the equation of state $P_X = -\rho_X/3$, thus the corresponding Hubble rate takes the simple form
\begin{equation}
 H(z) = H_0 \sqrt{\Omega_{\Lambda} + \Omega_X(1+z)^2 + \Omega_m(1+z)^3},
\end{equation}
with $\Omega_\Lambda = 1 - \Omega_{X} - \Omega_m$. The process of measurement indeed differs,
since $d_L$ has a different expressions for flat and non-flat cases. In general, the luminosity distance
equation is
\begin{eqnarray}
 d_L &=& \frac{(1+z)}{\sqrt{\Omega_{k}}} \sinh (\sqrt{\Omega_{k}} \chi(z)) \nonumber \\
 &\approx& (1+z) \left( \chi(z) + \frac{1}{3!}\Omega_{k} \chi(z)^3 \right)\,, \label{lumdistNonFlat}
\end{eqnarray}
with $\chi(z)$ the comoving distance to redshift $z$ given by Eq.~(\ref{kjjgjhg}).
For sufficiently small $\Omega_{k}$, the second equality in equation~(\ref{lumdistNonFlat}) is a good approximation. For illustration purposes,
we can consider $d_L = \chi(z)$ and assume that
the estimated values for the parameters are good for small $\Omega_{k}$ and definitively identify $\Omega_X$ with curvature.

The cosmographic parameters up to $s_0$ are in this case

\begin{eqnarray}
 q_0 &=& -1 + \Omega_X + \frac{3}{2}\Omega_m\,, \nonumber \\
 j_0 &=& 1-\Omega_X \,, \\
 s_0 &=& (1-\Omega_X)^2 - \frac{3}{2}(3 - \Omega_X)\Omega_m\,.  \nonumber
\end{eqnarray}

From the second equation and the results for the Pad\'e approximant $P_{31}$ (Eq.~(\ref{P31expectedv})), we have $\Omega_X(j_0) = -0.16 \pm 0.37$.
In the case of $\Omega_m$, using the first and second equations, $\Omega_m(q_0) = 0.37 \pm 0.31$,
and using the second and third equations, $\Omega_m(s_0) = 0.29 \pm 0.32$. Joining these results we obtain
\begin{eqnarray}
 \Omega_m &=& 0.32 \pm 0.22\,, \\
 \Omega_X &=& -0.16 \pm 0.37\,.
\end{eqnarray}

\subsection{The case of the Chevallier-Polarski-Linder parametrization}

The Chevallier-Polarski-Linder (CPL)~\cite{cpl} dark energy parametrization assumes that  the universe is composed by baryons,
cold dark matter and dark energy with an evolving equation of state of the form:
\begin{equation}
 w_{de} = w_0 + w_a\frac{z}{1+z}\,.
\end{equation}
The background cosmology cannot distinguish between dark matter and baryons, thus we write
\begin{eqnarray}
 H(z) & = & H_0 \Bigg(\Omega_m (1+z)^3  + \Omega_{de} (1+z)^{3(1+w_0+w_a)} \times\nonumber\\
 & &  \exp \left(-\frac{3w_a z}{1+z}\right) \Bigg)^{1/2}\,,  \label{CPLH}
\end{eqnarray}
where $\Omega_m = \Omega_b + \Omega_c$ and $\Omega_{de} = 1-\Omega_m$. Using~(\ref{CPLH}) and Eqs.~(\ref{Hpunto}) we obtain

\begin{equation}
q_0 = \frac{1}{2} \left(1 + 3 w_0 \left(1 - \Omega_m \right)\right)\,,
\end{equation}

\begin{equation}
j_0 = \frac{3}{2} \big( (3 w_0 \left(w_0+1\right) +  w_a )(1 -\Omega_m ) \big) +1\,,
\end{equation}

\begin{eqnarray}
s_0 &=& \frac{1}{4} \Big[ 9 w_0 (1-\Omega_m) (w_a
   (\Omega_m-7)-9) \nonumber\\ &-&   33 w_a (1 - \Omega_m)   -  27 w_0^3  (1-\Omega_m) (3-\Omega_m)
  \nonumber\\ & +&   9 w_0^2 (1 - \Omega_m) (3 \Omega_m-16) - 14 \Big]\,.
\end{eqnarray}
Thus,  if we use the $P_{31}$ results, we have to estimate 3 parameters out of other 3 parameters.
This is done numerically by  propagating errors in Eqs.~(\ref{P31expectedv}) obtaining

\begin{eqnarray}
 \Omega_m &=& 0.26 \pm 0.19\,, \nonumber\\
 w_0 &=& -1.04 \pm 0.16\,, \\
 w_a &=& 0.08 \pm 0.28\,. \nonumber
\end{eqnarray}

\subsection{The case of unified dark energy}

One relevant approach to dark energy suggests that  the universe is composed by a single fluid,
which unifies dark matter and dark energy in a single description.
A barotropic perfect fluid with vanishing adiabatic sound speed reproduces the  $\Lambda$CDM behavior at the background,
as proposed in~\cite{Luongo:2011yk} and it is compatible with small perturbations, as shown in~\cite{Aviles:2011ak}. The corresponding equation of state reads
\begin{equation}
 w_{df} = -\frac{1}{1+A (1+z)^3}\,,
\end{equation}
while the total equation of state for the $\Lambda + \text{dm}$ total dark fluid in the $\Lambda$CDM model reads
\begin{equation}
 w_{\Lambda + \text{dm}} =\frac{  \sum_{i}  \rho_i w_i}{\sum_{i} \rho_i} = -\frac{1}{1+\frac{\Omega_{\text{dm}}}{\Omega_{\Lambda}}(1+z)^3}\,. \label{TEOSLCDM}
\end{equation}
Thus, both models, i.e. $\Lambda$CDM and the negligible sound speed model,
exactly behave in the same way and hence they are degenerate. There are several other options for a unified dark fluid which does  not
degenerate with $\Lambda$CDM. One of these frameworks is represented by the Chaplygin
gas~\cite{Kamenshchik:2001cp,Bilic:2001cg} and its generalizations~\cite{Bento:2002ps,Aviles:2011ak,Aviles:2011sd}
and constant adiabatic speed of sound models~\cite{Xu:2011bp}, among others. Therefore, we parameterize the dark fluid equation of state by a Taylor series:
\begin{equation}
 w_{df}(z) = w_0 + w_1 z + w_2 z^2 + w_3 z^3 +\mathcal{O}(4)\,,
\end{equation}
where
\begin{equation}
 w_i = \frac{1}{i!}\frac{d^i w(z)}{dz^i}\Big|_{z=0}\,.
\end{equation}
Knowing the value of $\Omega_b$, in order to estimate up to $w_i$, we need to use the first $i$th cosmographic parameters.
If we use Pad\'e approximants up to  $P_{23}$, then we truncate the expansion series
at third order. The Hubble rate easily reads
\begin{equation}
 H(z) = H_0 \sqrt{\Omega_b(1+z)^3+ \Omega_{df}F(z)}\,\,,
\end{equation}
where $\Omega_{df} = 1-\Omega_{b}$ and we define
\begin{eqnarray}
 F(z) &=& (1+z)^{3(1+w_0-w_1+w_2-w_3)} \times \nonumber\\
 \!\!\!\!\!\ \!\!\!\!\! \!\!\!\!\! \!\!\!\!\! \!\!\!\!\! \!\!\!\!\! & & \!\!\!\!\! \!\!\!\!\! \!\!\!\!\! \!\!\!\!\! \!\!\!\!\! \!\!\!\!\!
 \exp \left[ 3(w_1-w_2+w_3)z  + \frac{3}{2}(w_2-w_3) z^2 + w_3 z^3\right]. \label{fofZ}
\end{eqnarray}
The parameters to estimate are given implicitly by
\begin{eqnarray}
 q_0 &=& \frac{1}{2} + \frac{2}{3}w_0(1-\Omega_b)\,,  \label{DFq0} \\
 j_0 &=& 1 + \frac{9}{2} w_0(1+w_0) + \frac{3}{2}w_1 \nonumber \\
     & & - \frac{3}{2}\Omega_b \big( 3w_0(1+w_0) + w_1 \big)\,, \label{DFj0} \\
 s_0 &=& -\frac{7}{2} - \frac{9}{4}w_0(1-\Omega_b)(9+w_1(7-\Omega_b)) \nonumber\\
 & & -\frac{9}{4} w_0^2(1-\Omega_b)(16-3\Omega_b) \nonumber\\
 & & -\frac{27}{4} w_0^3(3-\Omega_b)(1-\Omega_b) \nonumber\\
 & &- \frac{45}{2}w_1(1-\Omega_b) - 3w_2(1-\Omega_b)\,, \label{DFs0}
\end{eqnarray}
which reduce to the flat-$\Lambda$CDM values when $w_0 = -1$, $w_1=w_2=0$, and by considering $\Omega_b \rightarrow \Omega_m$.
From several independent observations we have measurements of the baryon species in the universe. In this section we will take the best-fit
from the Planck Collaboration $\Omega_b = 0.0488$~\cite{Ade:2013zuv}. We report the estimated values from Pad\'e approximants $P_{21}$,
$P_{31}$ and $P_{23}$:

\begin{enumerate}
 \item Pad\'e approximant $P_{21}$:
  \begin{eqnarray}
 w_0 &=& -0.67 \pm 0.05\,, \nonumber \\
 w_1 &=& 0.37 \pm 0.13\,.
\end{eqnarray}

 \item Pad\'e approximant $P_{31}$:
 \begin{eqnarray}
 w_0 &=& -0.77 \pm 0.07\,, \nonumber \\
 w_1 &=& 0.63 \pm 0.37\,, \\
 w_2 &=& 0.06 \pm 0.50\,. \nonumber
\end{eqnarray}

 \item Pad\'e approximant $P_{23}$:
 \begin{eqnarray}
 w_0 &=& -0.87 \pm 0.12\,, \nonumber \\
 w_1 &=& 1.13 \pm 1.09\,, \nonumber\\
 w_2 &=& 0.23 \pm 2.71\,,  \\
 w_3 &=& -0.95 \pm 2.21\,. \nonumber
\end{eqnarray}

\end{enumerate}

These results should be compared with the best fit values for the $\Lambda$CDM model, $w_0 = -0.76$, $w_1 = 0.55$, $w_2 = 0.15$ and
$w=-0.32$, obtained by substituting in equation~(\ref{TEOSLCDM}) the values $\Omega_m = 0.2880$, $\Omega_{\Lambda} = 0.7119$,
estimated in Sec.~\ref{ParameterEstimations} for the $\Lambda$CDM model, and the value $\Omega_b = 0.0488$ from Planck.

\section{The universe equation of state}\label{universeEoS}

Now, let us consider an arbitrary collection of fluids (baryons, cold dark matter, dark energy, ...) with total energy density
$\rho = \sum_i \rho_i$ which comprises all possible species present in the universe. The Friedmann equation is thus $3H^2 = 8\pi G \rho$, as already mentioned.
We want to estimate the total equation of state of the universe given by
\begin{equation}
 w_{T}(z) = w_0 + w_1 z + w_2 z^2 + w_3 z^3 + \mathcal{O}(4)\,. \label{EOSUwT}
\end{equation}
The Friedmann equation can be recast as
\begin{equation}
 H(z) = H_0 \sqrt{F(z)}\,,
\end{equation}
where $F(z)$ is given again by Eq.~(\ref{fofZ}). The cosmographic parameters are equal to Eqs~(\ref{DFq0})-(\ref{DFs0}), by imposing $\Omega_b=0$.
At late times, the total equation of state of the universe is given by
\begin{equation}
 w_{T} = -\frac{1}{1+\frac{\Omega_m}{\Omega_{\Lambda}}(1+z)^3}\,. \label{TEOSLCDM2}
\end{equation}

We report the estimated values from Pad\'e approximants $P_{21}$,
$P_{31}$ and $P_{23}$:

\begin{enumerate}
 \item Pad\'e approximant $P_{21}$:
  \begin{eqnarray}
 w_0 &=& -0.64 \pm 0.05\,, \nonumber \\
 w_1 &=& 0.41 \pm 0.12\,.
\end{eqnarray}

 \item Pad\'e approximant $P_{31}$:
\begin{eqnarray}
 w_0 &=& -0.73 \pm 0.07\,, \nonumber \\
 w_1 &=& 0.67 \pm 0.34\,, \\
 w_2 &=& -0.02 \pm 0.50\,. \nonumber
\end{eqnarray}

 \item Pad\'e approximant $P_{23}$:
 \begin{eqnarray}
 w_0 &=& -0.83 \pm 0.12\,, \nonumber \\
 w_1 &=& 1.17 \pm 1.02\,, \nonumber\\
 w_2 &=& 0.10 \pm 2.51\,,  \\
 w_3 &=& -0.47 \pm 1.93\,. \nonumber
\end{eqnarray}

\end{enumerate}

These results should be compared with the best fit values for the $\Lambda$CDM model, $w_0 = -0.71$, $w_1 = 0.62$, $w_2 = 0.08$ and
$w=-0.39$, obtained by substituting in equation~(\ref{TEOSLCDM2}) the values $\Omega_m = 0.2880$, $\Omega_{\Lambda} = 0.7119$,
estimated in Sec.~\ref{ParameterEstimations} for the $\Lambda$CDM model.

\begin{table*}
\caption{{\small Table of best fits and their likelihoods (1$\sigma$) for the redshifts functions $y_1$ and $y_4$, using the parameter sets
$\mathcal{B}$ and $\mathcal{C}$.}}

\begin{tabular}{c|c|c|c|c}

\hline\hline

{\small $\quad$ Parameter $\quad$}  &   {\small $\qquad$ $y_1 = z/(1+z)$ $\qquad$ }    &   {\small $\qquad$ $y_1 = z/(1+z)$ $\qquad$ }
                                    &   {\small  $\qquad$ $y_4 = \tan z$ $\qquad$}     &   {\small $\quad$ $y_4 = \tan z$ $\quad$}\\
{\small $\quad$           $\quad$}  &   {\small  set $\mathcal{B}$                }    &   {\small set $\mathcal{C}$   }
                                    &   {\small  set $\mathcal{B}$     }               &   {\small set $\mathcal{C}$}\\

\hline

{\small$H_0$}       & {\small $75.11$}{\tiny ${}_{-3.44}^{+3.29}$}        & {\small $73.17$}{\tiny ${}_{-3.38}^{+3.92}$}
                    & {\small $72.34$}{\tiny ${}_{-3.97}^{+3.55}$}        & {\small $72.58$}{\tiny ${}_{-4.31}^{+3.94}$}\\[0.8ex]

{\small$q_0$}       & {\small $-1.0642$}{\tiny ${}_{-0.1958}^{+0.2216}$}      & {\small $-0.8517$}{\tiny ${}_{-0.3695}^{+0.3795}$}
                    & {\small $-0.868$}{\tiny ${}_{-0.2763}^{+0.3165}$}      & {\small $-0.7501$}{\tiny ${}_{-0.3839}^{+0.3891}$}\\[0.8ex]

{\small$j_0$}       & {\small $2.991$}{\tiny ${}_{-1.109}^{+1.030}$}         & {\small $1.983$}{\tiny ${}_{-2.772}^{+2.646}$}
                    & {\small $2.142$}{\tiny ${}_{-1.448}^{+1.411}$}         & {\small $1.520$}{\tiny ${}_{-1.736}^{+2.123}$}\\[0.8ex]

{\small$s_0$}       & {\small $4.919$}{\tiny ${}_{-3.198}^{+3.909}$}         & {\small $1.591$}{\tiny ${}_{-6.469}^{+10.905}$}
                    & {\small $5.149$}{\tiny ${}_{-1.338}^{+2.210}$}         & {\small $-0.206$}{\tiny ${}_{-4.256}^{+4.960}$}\\[0.8ex]

{\small$l_0$}       & -- --                                                  & {\small $7.96$}{\tiny ${}_{-4.79}^{+47.83}$}
                    & -- --                                                  & {\small $-18.64$}{\tiny ${}_{-12.72}^{+21.60}$}\\[0.8ex]

\hline \hline

\end{tabular}

{\tiny Notes.
a. $H_0$ is given in Km/s/Mpc units.
}

\label{table:1DOtherRedshifts}
\end{table*}

\section{Consequences of the Pad\'e results for dark energy}\label{consequencespade}

We showed that the use of Pad\'e approximants in cosmography provides a new
model-independent technique for reconstructing the luminosity distance and the Hubble parameter $H(z)$.
This method is particulary valid since standard constructions in cosmography require to develop the luminosity distance $d_{L}$
as a Taylor series and then to match the data with this approximation.
In particular, when data are taken over $z>1$, Pad\'e functions work better than truncated Taylor series.
To make the argument consistent, we have performed in Sec.~\ref{Sect:DataSet} a detailed analysis of our models derived from Pad\'e approximants with respect to the
data taken from different observations.
The results have been elaborated in Secs.~\ref{Sect:DataSet} and~\ref{ParameterEstimations} and compared with
the standard cosmographic approach and to the values inferred from assuming the $\Lambda$CDM model.
As expected, not all the Pad\'e approximants work properly. For example, we have commented that one has to take special care of the possible spurious
divergences that may appear in $d_{L}$ when approximating with Pad\'e, due to the fact that such functions are rational functions.

Moreover, not all Pad\'e models can fit the data in the appropriate
way. Indeed, we have seen both theoretically and numerically that approximants
whose degrees of the numerator and of the denominator are similar seem to be preferred
(see Figs.~\ref{LCDMDL}-\ref{fig:PA} and Tabs.~\ref{table:1DResults1}-\ref{table:1DResults3}).
This fact suggests that the increase of the luminosity distance with $z$ has to be indeed slower than the one depicted by a Taylor approximation.
Interestingly, our numerical analysis has singled out the Pad\'e functions $P_{21}$, $P_{31}$ and $P_{23}$,
which are the ones that draw the best samples,
with narrowest dispersion (see Figs.~\ref{fig:1dim2ModelB}-\ref{fig:Pade32}).
As one can see from Tabs.~\ref{table:1DResults1}-\ref{table:1DResults3}, the best fit values and errors for the CS parameters
estimated using the approximants $P_{21}$, $P_{31}$ and $P_{23}$ are in good agreement with the SC results.
In particular, the approximant $P_{23}$ gives smaller relative errors than the corresponding SC analysis, thus suggesting that enlarging the approximation order, the
analysis by means of Pad\'e are increasingly more appropriate than the standard one.
The estimated values of the CS parameters, through the use of the Pad\'e approximants $P_{21}$, $P_{31}$ and $P_{23}$
seem to indicate that the value of $H_{0}$ is smaller than the one derived by means of the standard (Taylor) approximation. Our results therefore
agree  with Planck results, which show smaller values of $H_0$ than previous estimations. On the contrary,
$q_{0}$ seems to be larger than the result obtained by standard cosmography, while for $j_{0}$ the situation is less clear ($P_{21}$ and $P_{23}$ indicate
a smaller value, while $P_{31}$ a larger one). In any cases, the sign of $j_0$ is positive at a $68\%$ of confidence level. This fact, according to Sec. II,
provides a universe which starts decelerating at a particular redshift $z_{tr}$, named the transition redshift.

\noindent From the above considerations, a comparison of our results with the ones obtained previously using Pad\'e expansions is essential.
In particular, in \cite{OrlChri} the authors employed a $P_{12}$ Pad\'e approximant, motivating their choice by noticing that for $z\ll1$ the requirement $m>n$ could be
appropriate to describe the behavior of $d_L$.
Their idea was to propose this Pad\'e prototype and to use it for higher redshift domains. Their heuristic guess has not been compared in that work with respect to other $P_{nm}$ approximants.
Hence, the need of extending their approach has been achieved in the present paper, where we analyzed thoroughly which extensions work better.
Moreover, the authors adopted the $P_{12}$ Pad\'e approximant as a first example to describe the convergence radius in terms of the Pad\'e formalism,
providing discrepancies with respect to standard Taylor treatments.
Their numerical analyses were essentially based on SNeIa data only, while in our paper we adopted different data sets, i.e. baryonic acoustic oscillation,
Hubble space telescope measurements and differential age data, with improved numerical accuracies developed by using the CosmoMC code \cite{Lewis:2002ah,cosmomc_notes}.
As a consequence, we found that the cosmographic results obtained using $P_{21}$ are significantly different from the ones obtained using $P_{12}$.
Indeed, in \cite{OrlChri} the authors employed the $P_{12}$ approximant only, whereas in our paper we reported in Fig. 4 the plots of $P_{21}$,
which definitively provide the differences between $P_{21}$ and $P_{12}$.
In general, our results seem to be more accurate and general than the numerical outcomes of \cite{OrlChri}.
However,  we showed a positive jerk parameter, for sets $\mathcal A$ and $\mathcal B$, which is compatible with their results, albeit not strictly constrained to $j_0>1$, as they proposed.
Numerical outcomes for $H_0$ and $q_0$ lie in similar intervals with respect to \cite{OrlChri}.
Summing up, although the use of $P_{12}$ is possible \emph{a priori}, we demonstrated here that considering different models one can find
parameterizations that work better than $P_{12}$ and therefore are more natural candidates for further uses in upcoming works on cosmography.

Further, it is of special interest to look at the comparison of the numerical results obtained for the cases of $\Lambda$CDM, CPL and unified dark energy models by inserting
the values estimated by fitting the Pad\'e functions  $P_{21}$, $P_{31}$ and $P_{23}$ (see Sec.~\ref{applicationsPade}).
From this analysis it turns out that all of them suggest small departures
from $\Lambda$CDM, as expected. Moreover, $P_{31}$ is the one that better reproduces the results of $\Lambda$CDM. However, we expect from Fig.~\ref{LCDMDL}
that $P_{22}$ and $P_{32}$ should match even better the $\Lambda$CDM predictions. Therefore, we consider that it is needed to repeat the analysis with a larger set of data in the region $z\gg1$
to get more reliable results in this sense. This indication will be object of extensive future works.
Finally, let us comment the fact that results of Tab.~\ref{table:1DOtherRedshifts}, compared with the ones in Tabs.~\ref{table:1DResults1}-\ref{table:1DResults3}, show that
the approximants $P_{21}$, $P_{31}$ and $P_{23}$ give values for the CS parameters that are much closer to the ones estimated by standard cosmography
and by the $\Lambda$CDM model, than the results provided by the introduction of auxiliary variables in the standard cosmographic approach.
This definitively candidates Pad\'e approximants to represent a significative alternative to overcome the issues of divergence in cosmography, without the need of any additional auxiliary parametrization.

\section{Final outlooks and perspectives}\label{conclusions}

In this work we
proposed the use of Pad\'e approximations in the context of observational cosmology.
In particular, we improved the standard cosmographic approach,
{which enables to} accurately determine refined cosmographic bounds  {on the dynamical parameters of the models}.
We  stressed the fact that the Pad\'e recipe
 {can be used as a} relevant tool to extend standard Taylor treatments  {of the luminosity distance.}
Our main goal was to introduce a class of Pad\'e approximants able to overcome all the problems plaguing modern cosmography.
To do so, we enumerated the basic properties and the most important demands of the Pad\'e treatment and we matched theoretical predictions with modern data.
In particular, the  {main advantage} of the rational cosmographic method  {is that}  Pad\'e functions
 {reduce  the issue of convergence of the standard cosmographic approach based on} truncated Taylor series,  {especially for data taken over
a larger redshift range.}
In other words, usual model independent expansions performed at $z=0$ suffer from divergences due to data spanning cosmic intervals with $z>1$.
Since Pad\'e  {approximants are rational functions, thence they can} naturally overcome this issue. In particular, in our numerical treatment, we have considered all the possible Pad\'e approximants of the luminosity distance whose order
of the numerator and denominator sum up to three, four and five and compared them with the corresponding Taylor polynomials
of degree three, four and five in $z$. Among these models, it turned out that the Pad\'e technique can give results similar to those obtained by standard cosmography
and also improve the accuracy. In addition, the Pad\'e technique overcomes the need of introducing auxiliary variables,
as proposed in standard cosmography to reduce divergences at higher redshifts.
To do so, we compared Pad\'e results also with Taylor re-parameterized expansions.  {In all the cases considered here,
our Pad\'e numerical outcomes appear to improve} the standard analyses.

Furthermore, we also considered to overcome the degeneracy problem by   {employing} additional data sets. In particular, we assumed union 2.1  type Ia supernovae, baryonic acoustic oscillation, Hubble space telescope measurements and direct observations of Hubble rates,
based on the differential age method. Moreover, all cosmographic drawbacks have also been investigated and treated in terms of Pad\'e's recipe,
proposing for each problem a possible solution to improve the experimental analyses. Afterwards, we guaranteed our numerical outcomes to  {lie} in viable intervals and we demonstrated that the refined cosmographic bounds almost confirm the standard cosmological paradigm,
  {thus} forecasting the sign of the variation of acceleration, i.e. the jerk parameter.  However, although the $\Lambda$CDM model  {passes our experimental tests, we cannot conclude that} evolving dark energy terms are ruled out.
 Indeed, we compared our Pad\'e results with a class of cosmological models, namely
 the $\omega$CDM model, the Chevallier-Polarski-Linder parametrization and the unified dark energy models, finding a good agreement with those paradigms. Furthermore, we also investigated the consequences of Pad\'e's bounds on the universe equation of state.  {To conclude, we have proposed and investigated here the use of Pad\'e approximants in} the field of precision cosmology,
with particular regards to cosmography. Future perspectives will be clearly devoted to describe the Pad\'e approach in other relevant fields.
For example, early time cosmology is expected to be  {more easily} described in our framework, as well as additional epochs  {related to} high redshift data.  {Collecting all these results one could in principle definitively} reconstruct the universe expansion history, matching late with early time observations and also permitting to understand
whether the dark energy fluid evolves or remains a pure cosmological constant at all times.

\section*{Acknowledgements}

SC and OL are grateful to Manuel Scinta for important discussions on numerical and theoretical results. AB and OL want to thank Hernando Quevedo
and Christine Gruber for their support during the phases of this work.
AA is thankful to Jaime Klapp for discussions on the numerical outcomes of this work. SC is supported
by INFN through iniziative specifiche NA12, OG51.  OL is supported by the European PONa3 00038F1 KM3NET
(INFN) Project. AA is supported by the project CONACyT-EDOMEX-2011-C01-165873. AB  {wants to thank the} A. Della Riccia Foundation (Florence, Italy)  {for support.}

\begin{widetext}

\appendix
\section{Formulas used for approximating $d_{L}$ and OHD Table}\label{appA}
In this Appendix we give the formulas for the approximants of the luminosity distance used to fit the data, for every Taylor and Pad\'e approximant considered in this work.
Moreover, we provide also a Table of the Observational Hubble Data (OHD) used in the analysis.

The Taylor polynomials around $z=0$ of degree $3$, $4$ and $5$ for the luminosity distance~(\ref{defDL}) are given by
\begin{eqnarray}\label{Taylor1}
T3&=&
\frac{z}{6 H_{0}^3} \left[2 z^2 (H'_{0})^2-H_{0} z \left(z H''_{0}+3 (z+1) H'_{0}\right)+6 H_{0}^2 (z+1)\right]\,,\nonumber\\
\nonumber\\
T4&=&
	\frac{-z}{24 (H'_{0})^4}
	\Big[6 z^3 (H'_{0})^3-2 H_{0} z^2 H'_{0} \left(3 z H''_{0}+4 (z+1) H'_{0}\right)
	+H_{0}^2 \left(H^{(4)}_{0} z^3+4 (z+1) z (z H^{(3)}_{0}+3 H''_{0})\right)-24 H_{0}^3 (z+1)\Big]\,,\nonumber\\
\nonumber\\
T5&=&
	\frac{-z}{120 H_{0}^5} \Big\{-24 z^4 (H'_{0})^4+6 H_{0} z^3 (H'_{0})^2 \left(6 z H''_{0}+5 (z+1) H'_{0}\right)-2 H_{0}^2 z^2 \Big[3 z^2 (H''_{0})^2+20 (z+1) (H'_{0})^2\nonumber\\
	&+&z H'_{0} \left(4 H^{(3)}_{0} z+15 (z+1) H''_{0}\right)\Big]+H_{0}^3 \left[H^{(4)}_{0} z^4+5 (z+1) z \left(12 H'_{0}+z \left(H^{(3)}_{0} z+4 H''_{0}\right)\right)\right]-120 H_{0}^4 (z+1)\Big\}\,.\nonumber
\end{eqnarray}

Therefore, using Eqs.~(\ref{Hpunto}) and~(\ref{cosmoz}),
one can rewrite the Taylor approximations for the luminosity distance in terms of the CS parameters,
that are
\begin{eqnarray}\label{Taylor2}\nonumber
T3&=&
\frac{z}{6 H_{0}} \Big[z (-(j+1) z+q (3 q z+z-3)+3)+6\Big]\,,\nonumber\\
\nonumber\\
T4&=&\frac{z}{24 H_{0}} \left[z^3 (5 j (2 q+1)-q (3 q+2) (5 q+1)+s+2)-4 z^2 (j-q (3 q+1)+1)-12 (q-1) z+24\right]\,,\nonumber\\
\nonumber\\
T5&=&\frac{z}{120 H_{0}}
\Big[z^4 \left(10 j^2-j (5 q (21 q+22)+27)-l+q (q (q (105 q+149)+75)-15 s+6)-11 s-6\right)\nonumber\\
&+&5 z^3 (5 j (2 q+1)-q (3 q+2) (5 q+1)+s+2)-20 z^2 (j-q (3 q+1)+1)-60 (q-1) z+120\Big]\,,\nonumber
\end{eqnarray}\nonumber
where all the CS parameters are assumed to be evaluated at $z=0$.

At the same time, we can write all the Pad\'e approximants used in this work for the luminosity distance, which read
\begin{eqnarray}\label{Pade1}
P_{11}&=&{2  z}\left\{z H'_{0}-2 H_{0} (z-1)\right\}^{-1}\,,\nonumber\\
\nonumber\\
P_{12}&=&{12 H_{0} z}\left\{-z^2 (H'_{0})^2+2 H_{0} z \left(z H''_{0}-3 (z-1) H'_{0}\right)+12 H_{0}^2 ((z-1) z+1)\right\}^{-1}\,,\nonumber\\
\nonumber\\
P_{21}&=&{z \left(-z (H'_{0})^2+2 H_{0} \left(z H''_{0}-3 (z+1) H'_{0}\right)+12 H_{0}^2 (z+1)\right)}\left\{2 H_{0} \left(-2 z (H'_{0})^2+H_{0} \left(z H''_{0}+3 (z-1) H'_{0}\right)+6 H_{0}^2\right)\right\}^{-1}\,,\nonumber\\
\nonumber\\
P_{13}&=&-24 H_{0}^2 z
\Big\{-z^3 (H'_{0})^3+2 H_{0} z^2 H'_{0} \left(z H''_{0}-(z-1) H'_{0}\right)
-H_{0}^2 z \left(12 ((z-1) z+1) H'_{0}+z \left(H^{(3)}_{0} z-4 (z-1) H''_{0}\right)\right)\nonumber\\
&+&24 H_{0}^3 (z-1) \left(z^2+1\right)\Big\}^{-1}\,,\nonumber\\
\nonumber\\
P_{22}&=&H_{0}
\Big\{6 z \left(-z (H'_{0})^3-2 H_{0} H'_{0} \left((z+1) H'_{0}-z H''_{0}\right)+H_{0}^2 \left(-H^{3}_{0} z+4 (z+1) H''_{0}-12 (z+1) H'_{0}\right)+24 H_{0}^3 (z+1)\right)\Big\}\nonumber\\
&\,&
\Big\{-2 z^2 (H'_{0})^4+2 H_{0} z (H'_{0})^2 \left(z H''_{0}+6 (z-1) H'_{0}\right)+6 H_{0}^3 \left(H^{4}_{0} (z-1) z+4 \left(z^2+1\right) H''_{0}+12 (z-1) H'_{0}\right)\nonumber\\
&-&H_{0}^2 \left(-4 z^2 (H^{(3)}_{0})^2+12 (z (z+3)+1) (H'_{0})^2+3 z H''_{0} \left(H^{3}_{0} z+8 (z-1) H''_{0}\right)\right)+144 (H'_{0})^4\Big\}^{-1}\,,\nonumber\\
\nonumber\\
P_{31}&=&\Big\{z \Big[2 z^2 (H'_{0})^4-2 H_{0} z (H'_{0})^2 \left(z H''_{0}+6 (z+1) H'_{0}\right)\nonumber\\
&+&H_{0}^2 \left(-4 z^2 (H''_{0})^2+12 (z-4) (z+1) (H'_{0})^2+3 z H'_{0} \left(H^{(3)}_{0} z+8 (z+1) H''_{0}\right)\right)\nonumber\\
&-&6 H_{0}^3 (z+1) \left(H^{(3)}_{0} z+4 (z-1) H''_{0}-12 H'_{0}\right)\Big]\Big\}\nonumber\\
&\,&\Big\{6 H_{0}^2 \left(-6 z (H'_{0})^3+2 H_{0} H'_{0} \left(3 z H''_{0}+4 (z-1) H'_{0}\right)+H_{0}^2 \left(-H^{(3)}_{0} z-4 (z-1) H''_{0}+12 H'_{0}\right)\right)\Big\}^{-1}\,,\nonumber\\
\nonumber\\
P_{14}&=
&{720 H_{0}^3 z}
\Big\{-19 z^4 (H'_{0})^4+2 H_{0} z^3 (H'_{0})^2 \left(23 z H''_{0}-15 (z-1) H'_{0}\right)-2 H_{0}^2 z^2 \Big[8 z^2 (H''_{0})^2+30 ((z-1) z+1) (H'_{0})^2\nonumber\\
&+&3 z H'_{0} \left(3 H^{3}_{0} z-10 (z-1) H''_{0}\right)\Big]-6 H_{0}^3 z \left(-H^{4}_{0} z^3+60 (z-1) \left(z^2+1\right) H'_{0}+5 z \left(H^{3}_{0} (z-1) z-4 ((z-1) z+1) H''_{0}\right)\right)\nonumber\\
&+&720 H_{0}^4 \left((z-1) z \left(z^2+1\right)+1\right)\Big\}^{-1}\,,\nonumber\\
\nonumber\\
P_{41}&=&
\Big\{z \Big[-12 z^3 (H'_{0})^6+24 H_{0} z^2 (H'_{0})^4 \left(z H''_{0}+2 (z+1) H'_{0}\right)+24 H_{0}^5 (z+1) \left(H^{4}_{0} z+5 H^{3}_{0} (z-1)-20 H''_{0}\right)\nonumber\\
&-&4 H_{0}^2 z (H'_{0})^2 \left(3 z^2 (H''_{0})^2+2 (z+1) (5 z+27) (H'_{0})^2+z H'_{0} \left(H^{3}_{0} z+18 (z+1) H''_{0}\right)\right)\nonumber\\
&+&H_{0}^4 \Big[960 (z+1) (H'_{0})^2+z \left(5 (H^{(3)}_{0})^2 z^2+16 (z+1) (5 z-9) (H''_{0})^2+4 z \left(5 H^{3}_{0} (z+1)-H^{4}_{0} z\right) H''_{0}\right)\nonumber\\
&-&12 (z+1) H'_{0} \left(20 (2 z-3) H''_{0}+z \left(H^{4}_{0} z+H^{3}_{0} (5 z+11)\right)\right)\Big]+4 H_{0}^3 \Big(6 z^3 (H''_{0})^3+60 (z-3) (z+1) (H'_{0})^3\nonumber\\
&-&z^2 H'_{0} H''_{0} \left(7 H^{3}_{0} z+12 (z+1) H'_{0}\right)+2 (H'_{0})^2 \left(H^{4}_{0} z^3+(z+1) z \left(7 H''_{0} z+(5 z+63) H''_{0}\right)\right)\Big)\Big]\Big\}\nonumber\\
&\,&
\Big\{24 H_{0}^3 \Big[-24 z (H'_{0})^4+6 H_{0} (H'_{0})^2 \left(6 z H''_{0}+5 (z-1) H'_{0}\right)+H_{0}^3 \left(H^{3}_{0} z+5 H^{3}_{0} (z-1)-20 H''_{0}\right)\nonumber\\
&+&H_{0}^2 \left(-6 z (H''_{0})^2+40 (H'_{0})^2+2 H'_{0} \left(-4 H^{3}_{0} z-15 (z-1) H''_{0}\right)\right)\Big]\Big\}^{-1}\,,\nonumber\\
\nonumber\\
P_{32}&=&
\Big\{z \Big[2 z^2 (H'_{0})^4-2 H_{0} z (H'_{0})^2 \left(z H''_{0}+6 (z+1) H'_{0}\right)\nonumber\\
&+&H_{0}^2 \left(-4 z^2 (H''_{0})^2+12 (z-4) (z+1) (H'_{0})^2+3 z H'_{0} \left(H^{3}_{0} z+8 (z+1) H''_{0}\right)\right)\nonumber\\
&-&6 H_{0}^3 (z+1) \left(H^{3}_{0} z+4 (z-1) H''_{0}-12 H'_{0}\right)\Big]\Big\}
\Big\{6 H_{0}^2 \Big[-6 z (H'_{0})^3+2 H_{0} H'_{0} \left(3 z H''_{0}+4 (z-1) H'_{0}\right)\nonumber\\
&+&H_{0}^2 \left(-H^{3}_{0} z-4 (z-1) H''_{0}+12 H'_{0}\right)\Big]\Big\}^{-1}\,,\nonumber\\
\nonumber\\
P_{23}&=&
\Big\{6 z \left(-z (H'_{0})^3-2 H_{0} H'_{0} \left((z+1) H'_{0}-z H''_{0}\right)+H_{0}^2 \left(-H^{3}_{0} z+4 (z+1) H''_{0}-12 (z+1) H'_{0}\right)+24 H_{0}^3 (z+1)\right)\Big\}\nonumber\\
&\,&\Big\{-2 z^2 (H'_{0})^4+2 H_{0} z (H'_{0})^2 \left(z H''_{0}+6 (z-1) H'_{0}\right)+6 H_{0}^3 \left(H^{3}_{0} (z-1) z+4 \left(z^2+1\right) H''_{0}+12 (z-1) H'_{0}\right)\nonumber\\
&-&H_{0}^2 \left(-4 z^2 (H''_{0})^2+12 (z (z+3)+1) (H'_{0})^2+3 z H'_{0} \left(H^{3}_{0} z+8 (z-1) H''_{0}\right)\right)+144 H_{0}^4\Big\}^{-1}H^{4}_{0}\,.\nonumber
\end{eqnarray}
\\
\\
\indent Again, using Eqs.~(\ref{Hpunto}) and~(\ref{cosmoz}),
one can rewrite the Pad\'e approximants for the luminosity distance in terms of the CS parameters,
that are

\begin{eqnarray}\label{Pade1}
P_{11}&=& {2 z}\left\{H_{0} ((q-1) z+2)\right\}^{-1}\nonumber\,,
\\
\nonumber\\
P_{12}&=&
{-12 z}\left\{H_{0} (z (-(2 j+5) z+q ((3 q+8) z-6)+6)-12)\right\}^{-1}\nonumber\,,
\\
\nonumber\\
P_{21}&=&
{z (z (-2 j+q (3 q+8)-5)+6 (q-1))}\left\{2 H_{0} (-(j+1) z+q (3 q z+z+3)-3)\right\}^{-1}\nonumber\,,
\\
\nonumber\\
P_{13}&=&
{24 z}\left\{H_{0} \left(z^3 (-(j (6 q+9)-q (6 q (q+2)+19)+s+9))+2 z^2 (2 j-q (3 q+8)+5)+12 (q-1) z+24\right)\right\}^{-1}\nonumber\,,
\\
\nonumber\\
P_{22}&=&
\Big\{6 z (z (j (6 q+9)-q (6 q (q+2)+19)+s+9)+2 (2 j-q (3 q+8)+5))\Big\}
\Big\{H_{0} \Big[z^2 \Big(4 j^2+j (q (6 q-23)-7)\nonumber\\
&+&q \left(q \left(-9 q^2+30 q+13\right)+3 s+4\Big)-3 s-2\right)+6 z (j (8 q+7)-q (q (9 q+17)+6)+s+4)\nonumber\\
&+&12 (2 j-q (3 q+8)+5)\Big]\Big\}^{-1}\nonumber\,,
\\
\nonumber\\
P_{31}&=&
\Big\{z \Big[z^2 \left(-4 j^2+j (q (23-6 q)+7)+q \left(q \left(9 q^2-30 q-13\right)-3 s-4\right)+3 s+2\right)+6 z (j (8 q+7)\nonumber\\
&-&q (q (9 q+17)+6)+s+4)+24 (j-q (3 q+1)+1)\Big]\Big\}\Big\{6 H_{0} (z (5 j (2 q+1)-q (3 q+2) (5 q+1)+s+2)\nonumber\\
&+&4 (j-q (3 q+1)+1))\Big\}^{-1}\nonumber\,,
\\
\nonumber\\
P_{14}&=&
-720 z
\Big\{H_{0} \Big[z^4 \left(40 j^2-2 j (5 q (30 q+59)+221)-6 l+q (q (3 q (75 q+188)+610)-60 s+646)-96 s-251\right)\nonumber\\
&+&30 z^3 (j (6 q+9)-q (6 q (q+2)+19)+s+9)+60 z^2 (-2 j+q (3 q+8)-5)-360 (q-1) z-720\Big]\Big\}^{-1}\nonumber\,,
\\
\nonumber\\
P_{41}&=&
\Big\{z \Big[4 z^2 \Big(5 j^2 (4 q+11)+j (q (5 q (18 q-35)-234)+5 s-46)+3 l (q-1)+q \big(2 q \left(q \left(-45 q^2+69 q+121\right)+15 s+61\right)\nonumber\\
&-&17 s+16\big)-4 (7 s+2)\Big)+12 z \left(20 j^2-j (5 q (32 q+49)+79)-2 l+q (q (q (135 q+308)+205)-25 s+32)-27 s-22\right)\nonumber\\
&+&z^3 \big[-\big(40 j^3+j^2 (20 q (1-2 q)+57)+j (-4 l+2 q (q (q (90 q+143)-103)+4 (5 s-26))+6 s-32)-4 (l-2 q+6 s+1)\nonumber\\
&+&q (4 l (3 q+1)+q (q (184-3 q (q (45 q+86)-23))+108))+2 q (q (15 q+31)-18) s+5 s^2\big)\big]\nonumber\\
&-&120 (5 j (2 q+1)-q (3 q+2) (5 q+1)+s+2)\Big]\Big\}
\Big\{24 H_{0} \Big[z \Big(10 j^2-j (5 q (21 q+22)+27)-l+q (q (q (105 q+149)+75)\nonumber\\
&-&15 s+6)-11 s-6\Big)-5 (5 j (2 q+1)-q (3 q+2) (5 q+1)+s+2)\Big]\Big\}^{-1}\nonumber\,,
\\
\nonumber\\
P_{32}&=&
\Big\{z \Big[z^2 \left(-4 j^2+j (q (23-6 q)+7)+q \left(q \left(9 q^2-30 q-13\right)-3 s-4\right)+3 s+2\right)+6 z (j (8 q+7)\nonumber\\
&-&q (q (9 q+17)+6)+s+4)+24 (j-q (3 q+1)+1)\Big]\Big\}
\Big\{6 H_{0} (z (5 j (2 q+1)-q (3 q+2) (5 q+1)+s+2)\nonumber\\
&+&4 (j-q (3 q+1)+1))\Big\}^{-1}\nonumber\,,
\\
\nonumber\\
P_{23}&=&
\Big\{6 z (z (j (6 q+9)-q (6 q (q+2)+19)+s+9)+2 (2 j-q (3 q+8)+5))\Big\}
\Big\{H_{0} \Big[z^2 \Big(4 j^2+j (q (6 q-23)-7)\nonumber\\
&+&q \left(q \left(-9 q^2+30 q+13\right)+3 s+4\right)-3 s-2\Big)+6 z (j (8 q+7)-q (q (9 q+17)+6)+s+4)+12 (2 j-q (3 q+8)+5)\Big]\Big\}^{-1}\,,\nonumber
\end{eqnarray}
\\
where all the CS parameters are assumed to be evaluated at $z=0$.

For completeness, we include also here the Table of the OHD used in this work. They are summarized
in Table \ref{table:OHD}.

\begin{table*}
\caption{{\small Summary of OHD used in this paper. The top panel data use  passively evolving galaxies as cosmic chronometers; the bottom
panel uses data inferred from the study of different galaxy surveys. The standard deviations include model independent statistical
estimation error and systematics.}}

\begin{tabular}{c|c|c|c}

\hline\hline

{\small $\qquad z \qquad$}  &   {\small $\qquad H(z)\,^a \qquad$}                  & {\small $\qquad \sigma_H\,^a \qquad $}   & {\small $\quad$ Reference  $\quad$} \\

\hline
{\small$0.090$}       & {\small $69$}          & {\small $12$} & {\small  \cite{Simon:2004tf}} \\[0.2ex]
{\small$0.170$}       & {\small $83$}          & {\small $8$} & {\small  \cite{Simon:2004tf}} \\[0.2ex]
{\small$0.1791$}       & {\small $75$}          & {\small $4$} & {\small  \cite{Moresco:2012jh}} \\[0.2ex]
{\small$0.1993$}       & {\small $75$}          & {\small $5$} & {\small  \cite{Moresco:2012jh}} \\[0.2ex]
{\small$0.270$}       & {\small $77$}          & {\small $14$} & {\small  \cite{Simon:2004tf}} \\[0.2ex]

{\small$0.3519$}       & {\small $83$}          & {\small $14$} & {\small  \cite{Moresco:2012jh}} \\[0.2ex]

{\small$0.400$}       & {\small $95$}          & {\small $17$} & {\small  \cite{Simon:2004tf}} \\[0.2ex]

{\small$0.480$}       & {\small $97$}          & {\small $62$} & {\small  \cite{Stern:2009ep}} \\[0.2ex]

{\small$0.5929$}       & {\small $104$}          & {\small $13$} & {\small  \cite{Moresco:2012jh}} \\[0.2ex]
{\small$0.6797$}       & {\small $92$}          & {\small $8$} & {\small  \cite{Moresco:2012jh}} \\[0.2ex]
{\small$0.7812$}       & {\small $105$}          & {\small $12$} & {\small  \cite{Moresco:2012jh}} \\[0.2ex]
{\small$0.8754$}       & {\small $125$}          & {\small $17$} & {\small  \cite{Moresco:2012jh}} \\[0.2ex]

{\small$0.880$}       & {\small $90$}          & {\small $40$} & {\small  \cite{Stern:2009ep}} \\[0.2ex]

{\small$0.900$}       & {\small $117$}          & {\small $23$} & {\small  \cite{Simon:2004tf}} \\[0.2ex]

{\small$1.037$}       & {\small $154$}          & {\small $20$} & {\small  \cite{Moresco:2012jh}} \\[0.2ex]

{\small$1.300$}       & {\small $168$}          & {\small $17$} & {\small  \cite{Simon:2004tf}} \\[0.2ex]
{\small$1.430$}       & {\small $177$}          & {\small $18$} & {\small  \cite{Simon:2004tf}} \\[0.2ex]
{\small$1.530$}       & {\small $140$}          & {\small $14$} & {\small  \cite{Simon:2004tf}} \\[0.2ex]
{\small$1.750$}       & {\small $202$}          & {\small $40$} & {\small  \cite{Simon:2004tf}} \\[0.2ex]

\hline

{\small$0.2$}       & {\small $71$}          & {\small $8$} & {\small  \cite{Scrimgeour:2012wt}} \\[0.2ex]
{\small$0.24$}      & {\small $76.69$}      & {\small $3.61$} & {\small  \cite{Gaztanaga:2008xz}} \\[0.2ex]
{\small$0.4$}       & {\small $70$}          & {\small $5$} & {\small  \cite{Scrimgeour:2012wt}} \\[0.2ex]
{\small$0.43$}      & {\small $86.45$}      & {\small $4.96$} & {\small  \cite{Gaztanaga:2008xz}} \\[0.2ex]
{\small$0.6$}       & {\small $81$}          & {\small $5$} & {\small  \cite{Scrimgeour:2012wt}} \\[0.2ex]
{\small$0.8$}       & {\small $75$}          & {\small $4$} & {\small  \cite{Scrimgeour:2012wt}} \\[0.2ex]
{\small$2.3$}       & {\small $224$}          & {\small $8$} & {\small  \cite{Busca:2012bu}} \\[0.2ex]

\hline \hline

\end{tabular}

{\tiny Notes.
a. $H(z)$ and $\sigma_H$ are given in Km/s/Mpc units.
}

\label{table:OHD}
\end{table*}

To conclude, we also include here all the approximations for
the functions $H(z)$ and $D_{V}$ corresponding to the Pad\'e approximations $P_{nm}$ for $d_{L}$ used in this paper up to order $m+n=4$,
 following the prescription indicated in Sec. \ref{subsec:BAO}.
Starting from the expressions above and using Eq. (\ref{HfromDL}), one obtains the corresponding functions $H(z)$ as
\begin{eqnarray}\label{Pade1}
H_{11}&=& -\frac{(z+1)^2 ((q-1) z+2)^2 H_{0}}{2 (q-1) z^2-4}\nonumber\,,
\\
\nonumber\\
H_{12}&=&-\frac{((1 + z)^2 (12 + 6 (-1 + q) z + (5 + 2 j - q (8 + 3 q)) z^2)^2 H_{0}}{12 (-12 +
    z^2 (-1 + 10 z + j (2 + 4 z) - q (2 + 16 z + q (3 + 6 z)))))}
\nonumber\,,
\\
\nonumber\\
H_{21}&=&
\frac{2 (1 + z)^2 (3 + z + j z - q (3 + z + 3 q z))^2 H_{0}}{
18 (-1 + q)^2 +
 6 (-1 + q) (-5 - 2 j + q (8 + 3 q)) z + (14 + 2 j^2 +
  j (7 - q (10 + 9 q)) + q (-40 + q (17 + 9 q (2 + q)))) z^2}
\nonumber\,,\\
H_{13}&=&
\Big[(1 + z)^2 (24 + 12 (-1 + q) z +
   2 (5 + 2 j - q (8 + 3 q)) z^2 - (9 + j (9 + 6 q)
  -q (19 + 6 q (2 + q)) + s) z^3)^2 H_{0}\Big]  \Big[24 (24  \nonumber\\
  &+& z^2 (2 - 4 j + 4 q + 6 q^2 +
      2 (-1 + j (5 + 6 q) - 3 q (1 + 2 q (1 + q)) + s) z +
      3 (9 + j (9 + 6 q) - q (19 + 6 q (2 + q)) + s) z^2))\Big]^{-1}
\nonumber\,,
\\
\nonumber\\
H_{22}&=&
-\Big[(1 + z)^2 (12 (5 + 2 j - q (8 + 3 q)) +
      6 (4 + j (7 + 8 q) - q (6 + q (17 + 9 q)) + s) z + (-2 + 4 j^2 +
          j (-7 + q (-23 + 6 q)) \nonumber\\
          &-& 3 s +
         q (4 + q (13 + 30 q - 9 q^2) + 3 s)) z^2)^2 H_{0}\Big]
     \Big[6 (-24 (5 + 2 j - q (8 + 3 q))^2 -
      24 (5 + 2 j - q (8 + 3 q)) (9 + j (9 + 6 q) \nonumber\\
      &-&
         q (19 + 6 q (2 + q)) + s) z +
      2 (-268 + 8 j^3 - 9 j^2 (23 + 4 q (11 + 4 q)) -
         q (-1056 + q (384 + q (920 + 27 q (49 + q (22 + 5 q)))))\nonumber\\
         &+& 6 j (-76 + q (89 + q (236 + 9 q (18 + 5 q)) - 6 s) - 9 s) -
         54 s + 6 q (19 + 6 q (2 + q)) s - 3 s^2) z^2 \nonumber\\
      &+& 4 (5 + 2 j  - q (8 + 3 q)) (-2 + 4 j^2 +
         j (-7 + q (-23 + 6 q)) - 3 s +
         q (4 + q (13 + 30 q - 9 q^2) + 3 s)) z^3\nonumber\\
          &+& (9 + j (9 + 6 q)
         -q (19 + 6 q (2 + q)) + s) (-2 + 4 j^2 +
         j (-7 + q (-23 + 6 q)) - 3 s +
         q (4 + q (13 + 30 q - 9 q^2) + 3 s)) z^4)\Big]^{-1}
\nonumber\,,
\\
\nonumber\\
H_{31}&=&
-\Big[6 (1 + z)^2 (4 (1 + j - q (1 + 3 q)) + (2 + 5 j (1 + 2 q) -
         q (2 + 3 q) (1 + 5 q) + s) z)^2 H_{0}\Big]
         \Big[-96 (1 + j - q (1 + 3 q))^2 \nonumber\\
    &-& 48 (1 + j - q (1 + 3 q)) (4 + j (7 + 8 q) - q (6 + q (17 + 9 q)) +
        s) z + 6 (8 j^3 - j^2 (49 + 4 q (39 + 23 q)) \nonumber\\
      &-& q (-56 + q (-128 + q (112 + q (509 + 462 q + 81 q^2)))) +
       2 j (-34 + q (-2 + q (205 + q (281 + 78 q)) - 6 s) - 9 s) \nonumber\\
      &+& 2 q (10 + 3 q (7 + q)) s - s^2 - 4 (5 + 3 s)) z^2 +
    2 (6 + j (9 + 10 q) - q (6 + 5 q (5 + 3 q)) + s) (-2 + 4 j^2 +
       j (-7 + q (-23 + 6 q)) \nonumber\\
       &-& 3 s +
       q (4 + q (13 + 30 q - 9 q^2) + 3 s)) z^3 + (2 + 5 j (1 + 2 q) -
        q (2 + 3 q) (1 + 5 q) + s) (-2 + 4 j^2 +
       j (-7 + q (-23 + 6 q)) \nonumber\\
       &-& 3 s +
       q (4 + q (13 + 30 q - 9 q^2) + 3 s)) z^4\Big]^{-1}
       \nonumber\,,
\end{eqnarray}
where all the cosmographic parameters have been evaluated at $z=0$.

Afterwards, according to Eq. (\ref{defDV}), the corresponding $D_{V}$ functions are also reported:
\begin{eqnarray}
(D_{V})_{11}&=&
2 \left(\frac{2 z^3 - (-1 + q) z^5}{(1 + z)^4 (2 + (-1 + q) z)^4 H_{0}^3}\right)^{1/3}
\nonumber\,,
\\
\nonumber
(D_{V})_{12}&=&
12 \left(\frac{12 z^3 + (1 - 2 j + q (2 + 3 q)) z^5 +
   2 (-5 - 2 j + q (8 + 3 q)) z^6}
   {(1 + z)^4 (-12 +
     z (6 - (5 + 2 j) z + q (-6 + (8 + 3 q) z)))^4 H_{0}^3}\right)^{1/3}
\nonumber\,,
\\
\nonumber
(D_{V})_{21}&=&
\frac{1}{2}
\Big(\Big[z^3 (z (-2 j+q (3 q+8)-5)+6 (q-1))^2 \big(z^2 \left(2 j^2+j (7-q (9 q+10))+q (q (9 q (q+2)+17)-40)+14\right)\nonumber\\
&+& 6 (q-1) z (-2 j+q (3 q+8)-5)+18 (q-1)^2\big)\Big]\Big[H_0^3 (z+1)^4 (j z-q (3 q z+z+3)+z+3)^4\Big]^{-1}\Big)^{1/3}
\nonumber\,,
\\
\nonumber
(D_{V})_{13}&=&
24 \Big[\Big(z^3 (24 +
     z^2 (2 - 4 j + 4 q + 6 q^2 +
        2 (-1 + j (5 + 6 q) - 3 q (1 + 2 q (1 + q)) + s) z +
        3 (9 + j (9 + 6 q)\nonumber\\
        &-& q (19 + 6 q (2 + q)) + s) z^2))\Big)
        \Big((1 +
     z)^4 (24 + 12 (-1 + q) z +
     2 (5 + 2 j - q (8 + 3 q)) z^2 - (9 + j (9 + 6 q) \nonumber\\
      &-&  q (19 + 6 q (2 + q)) + s) z^3)^4 H_{0}^3\Big)^{-1}\Big]^{1/3}
\nonumber\,,
\\
\nonumber
(D_{V})_{22}&=&
-6 \Big[\Big(z^3 (2 (5 + 2 j - q (8 + 3 q)) + (9 + j (9 + 6 q) -
           q (19 + 6 q (2 + q)) + s) z)^2 (-24 (5 + 2 j -
           q (8 + 3 q))^2 \nonumber\\
       &-& 24 (5 + 2 j - q (8 + 3 q)) (9 + j (9 + 6 q) -
           q (19 + 6 q (2 + q)) + s) z +
        2 (-268 + 8 j^3 - 9 j^2 (23 + 4 q (11 + 4 q)) \nonumber\\
        &-&   q (-1056 + q (384 + q (920 + 27 q (49 + q (22 + 5 q))))) +
           6 j (-76 + q (89 + q (236 + 9 q (18 + 5 q)) - 6 s) -
              9 s)\nonumber\\
              &-& 54 s + 6 q (19 + 6 q (2 + q)) s - 3 s^2) z^2 +
        4 (5 + 2 j - q (8 + 3 q)) (-2 + 4 j^2 +
           j (-7 + q (-23 + 6 q)) - 3 s \nonumber\\
        &+&   q (4 + q (13 + 30 q - 9 q^2) + 3 s)) z^3 + (9 +
           j (9 + 6 q) - q (19 + 6 q (2 + q)) + s) (-2 + 4 j^2 +
           j (-7 + q (-23 + 6 q)) \nonumber\\
           &-& 3 s +   q (4 + q (13 + 30 q - 9 q^2) + 3 s)) z^4)\Big)
           \Big((1 +
        z)^4 (12 (5 + 2 j - q (8 + 3 q)) +
        6 (4 + j (7 + 8 q) \nonumber\\
        &-& q (6 + q (17 + 9 q)) + s) z + (-2 +
           4 j^2 + j (-7 + q (-23 + 6 q)) - 3 s +
           q (4 + q (13 + 30 q - 9 q^2) + 3 s)) z^2)^4 H_{0}^3\Big)^{-1}\Big]^{1/3}
\nonumber\,,
\\
\nonumber
(D_{V})_{31}&=&
-\frac{1}{6} \Big[\Big(z^3 (24 (1 + j - q (1 + 3 q)) +
        6 (4 + j (7 + 8 q) - q (6 + q (17 + 9 q)) + s) z - (-2 +
           4 j^2 + j (-7 + q (-23 + 6 q))\nonumber\\
           & -& 3 s +
           q (4 + q (13 + 30 q - 9 q^2) + 3 s)) z^2)^2 (-96 (1 + j -
           q (1 + 3 q))^2 -
        48 (1 + j - q (1 + 3 q)) (4 + j (7 + 8 q) \nonumber\\
        &-&   q (6 + q (17 + 9 q)) + s) z +
        6 (8 j^3 - j^2 (49 + 4 q (39 + 23 q)) -
           q (-56 + q (-128 + q (112 + q (509 + 462 q + 81 q^2))))\nonumber\\
        &+&   2 j (-34 + q (-2 + q (205 + q (281 + 78 q)) - 6 s) -
              9 s) + 2 q (10 + 3 q (7 + q)) s - s^2 -
           4 (5 + 3 s)) z^2 \nonumber\\
      &+&  2 (6 + j (9 + 10 q) - q (6 + 5 q (5 + 3 q)) + s) (-2 + 4 j^2 +
            j (-7 + q (-23 + 6 q)) - 3 s \nonumber\\
         &+&  q (4 + q (13 + 30 q - 9 q^2) + 3 s)) z^3 + (2 +
           5 j (1 + 2 q) - q (2 + 3 q) (1 + 5 q) + s) (-2 + 4 j^2 +
           j (-7 + q (-23 + 6 q)) \nonumber\\
           &-& 3 s +
           q (4 + q (13 + 30 q - 9 q^2) + 3 s)) z^4)\Big)
           \Big((1 +
        z)^4 (4 (1 + j - q (1 + 3 q)) + (2 + 5 j (1 + 2 q) \nonumber\\
          &-& q (2 + 3 q) (1 + 5 q) + s) z)^4 H_{0}^3\Big)^{-1}\Big]^{1/3}
\nonumber\,,
\\
\nonumber
\end{eqnarray}

\end{widetext}

\end{document}